\documentclass[12pt]{article}

\newenvironment{wileykeywords}{\textsf{Keywords:}\hspace{\stretch{1}}}{\hspace{\stretch{1}}\rule{1ex}{1ex}}

\usepackage{amsmath,amssymb}
\usepackage{graphicx}
\usepackage{color}
\usepackage{dcolumn}
\usepackage{bm}
\usepackage[numbers,square,super,comma,sort&compress]{natbib}

\usepackage{braket}
\usepackage{mathrsfs}
\usepackage{setspace}
\usepackage{mathptmx}
\usepackage{float}
\usepackage{booktabs}
\usepackage{csvsimple}
\usepackage{xcolor}

\definecolor{background-color}{gray}{0.98}

\title{Fitting continuum wavefunctions with complex
	Gaussians: Computation of ionization cross sections}
\author{A. Ammar, A. Leclerc and L.U.
	Ancarani\thanks{Universit\'e de Lorraine-CNRS, UMR 7019, LPCT, F-57000, France.} }

\begin{document}
	
	\newcommand{\tj}[6]{ \begin{pmatrix}
			#1 & #2 & #3 \\
			#4 & #5 & #6
	\end{pmatrix}}
	
	\maketitle
	
	\begin{abstract}
		We implement a full nonlinear optimization method to fit continuum
		states with complex Gaussians. The application to a set of regular
		scattering Coulomb functions allows us to validate the numerical
		feasibility, to explore the range of convergence of the approach,
		and to demonstrate the relative superiority of complex over real
		Gaussian expansions. We then consider the photoionization of atomic hydrogen, and ionization by electron impact in the first Born approximation, for which the closed form cross sections serve 
		as a solid benchmark.
		Using the proposed complex Gaussian representation of
		the continuum combined with a real Gaussian expansion for the
		initial bound state, all necessary matrix elements within a
		partial wave approach become analytical. The successful
		numerical comparison illustrates that the proposed all-Gaussian
		approach works efficiently for ionization processes of one-center
		targets.
	\end{abstract}
	
	\begin{wileykeywords}
		Continuum wavefunctions, Real Gaussians, Complex Gaussians, Non-linear optimization,
		Ionization.
	\end{wileykeywords}
	
	\clearpage
	
	
	\begin{figure}[h]
		\centering
		\colorbox{background-color}{
			\fbox{
				\begin{minipage}{1.0\textwidth}
					\begin{minipage}{0.5\textwidth}
						\includegraphics[width=\linewidth]{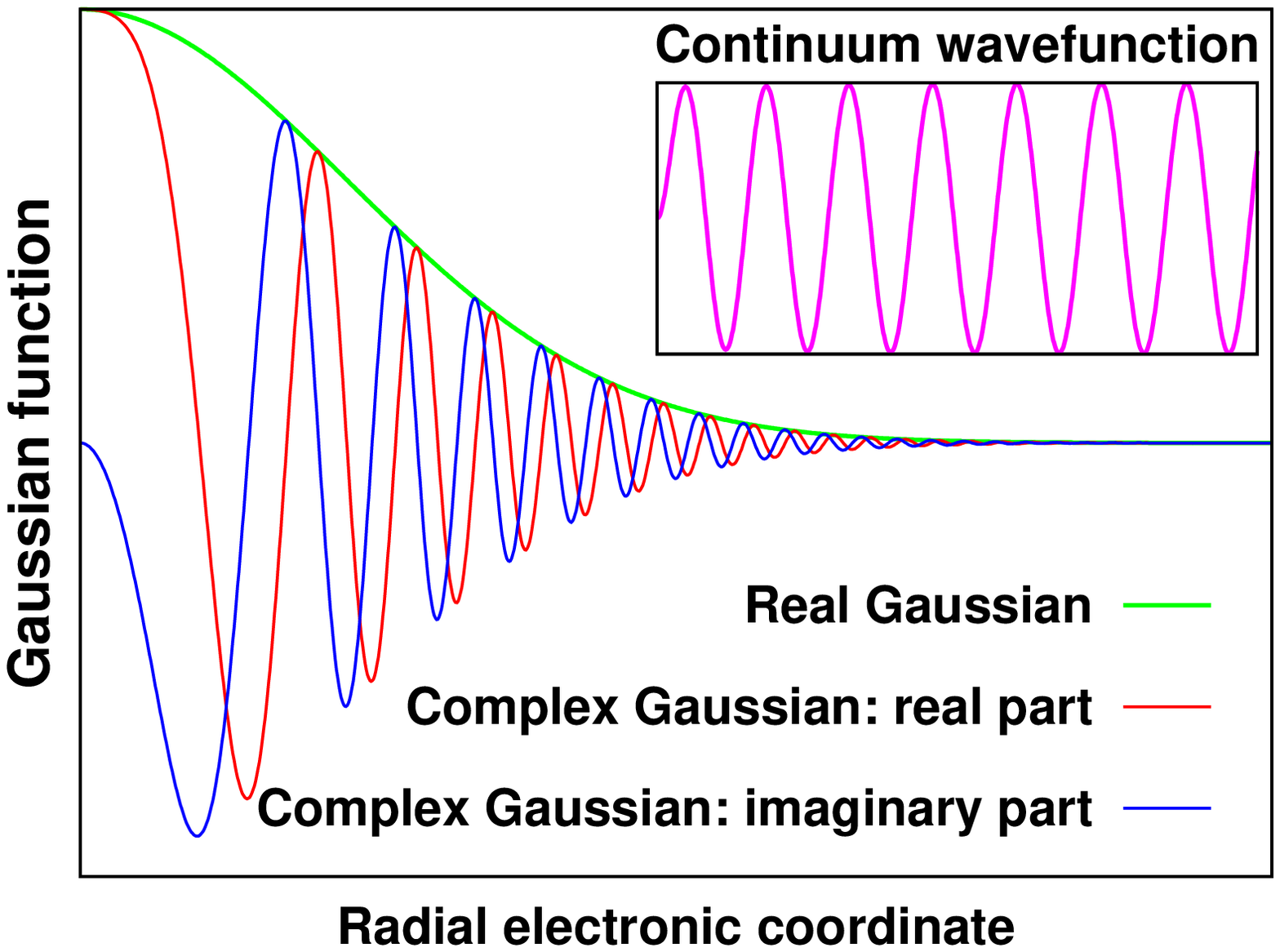}
					\end{minipage}
					\begin{minipage}{0.5\textwidth}
						We propose an all-Gaussian approach for electronic scattering
						processes with one-center targets. Gaussian functions are used not only to describe
						the initial bound states, but also to represent the final continuum states involved
						in processes such as photoionization or electron impact ionization. An efficient
						nonlinear optimization is performed to fit continuum wavefunctions with complex
						Gaussian representations and once the fitting is done, all integrals necessary
						to compute cross-sections become analytical.
					\end{minipage}
				\end{minipage}
		}}
	\end{figure}
	
	\makeatletter
	\renewcommand\@biblabel[1]{#1.}
	\makeatother
	
	\bibliographystyle{apsrev}
	
	\renewcommand{\baselinestretch}{1.5}
	\normalsize
	
	\clearpage
	
	\section{\sffamily \Large Introduction} 

	As early as $1950$ Boys~\cite{boys} emphasized the
	importance of Gaussians to simplify the calculation of multicenter
	integrals arising in the study of polyatomic molecular electronic
	states. One key property is the notorious Gaussian product
	theorem. For an overview of the use of Gaussian sets in molecular
	calculations see, \textit{e.g.}, reference~\cite{Hill}. The optimal
	set depends very much on the studied problem, \textit{e.g.},
	system energy minimization, calculation of resonance states,\dots.
	Usually, to avoid a full nonlinear optimization, one may consider
	a particular choice of exponents such as even-tempered
	sets~\cite{Reeves,Feller1979}, well-tempered
	sets~\cite{Huzinaga1}, random-tempered
	sets~\cite{Alexander1986} or polynomial
	expansions~\cite{Petersson}. In the $60$s many studies
	have been dedicated to the representation of bound states by a
	real Gaussian
	combination~\cite{Huzinaga1965,Reeves1965,oohata,Lim1966,stewart1969,stewart1970}. Despite their too fast decay at large distances, Gaussians
	manage to rather well reproduce molecular electronic orbitals over
	the physically relevant radial regions~\cite{Harris}.
	Another drawback is their mathematical inability to reproduce the
	correct cusp near the origin, an issue that is generally addressed
	numerically using a combination of Gaussians with large
	exponents~\cite{Huzinaga1965}.
	
	While Gaussians are pretty good for reproducing low-lying bound
	states, representing on large radial domains multinode
	states~(highly excited or Rydberg states), or highly oscillating
	continuum states by nodeless Gaussians is numerically very
	challenging. In spite of the numerous potential applications~(such
	as
	photoionization~\cite{Marante,Cacelli1,Cacelli2,Cacelli3},
	high harmonic generation~\cite{Coccia1,Coccia2,Coccia3},
	ionization by particle impact~\cite{Rmatrix}) only few
	exploratory studies have been
	made~\cite{Kaufmann_1989,nestmann,faure,fiori}. Kaufmann
	\textit{et al}~\cite{Kaufmann_1989} optimized a set of
	Gaussians to represent continuum functions with a reasonable
	accuracy up to $0.2~a.u.$ by the diagonalization of the attractive
	Coulomb Hamiltonian represented in a finite set of Gaussian-type
	functions. Nestmann and Peyerimhoff~\cite{nestmann}
	proposed a least square approach to represent Bessel functions
	with real Gaussians. Faure \textit{et al}~\cite{faure}
	applied the same approach to fit Bessel and Coulomb functions.
	However, the fitting error for large energies~(typically $\approx
	3$ Rydbergs) is visible to the naked eye. Fiori and
	Miraglia~\cite{fiori} optimized Gaussians to fit
	distortion  functions by first removing the dominant fast
	oscillatory plane wave from the Coulomb function. An alternative
	approach to reproduce the oscillating behavior of continuum states
	is to add to Gaussian sets some supplementary functions, like
	B-splines~\cite{Marante}.
	
	We believe that expansions in complex Gaussians, that it
	to say Gaussians with complex exponents, offer an alternative and
	possibly more suitable way of representing continuum states. 
	Complex Gaussians have already been proven to be successful in resonance
	stabilization calculations, where they arise from a complex scaling
	transformation~\cite{mccurdy1978,mccurdy1982,isaacson1991,white2017}. 
	Such functions possess an intrinsically oscillatory behavior that
	may facilitate the task of representing continuum states over larger radial
	domains. Previous
	investigation in this line, include the work of
	Matsuzaki \textit{et al} who used complex Slater
	orbitals~\cite{Matsuzaki1} and complex
	Gaussians~\cite{Matsuzaki2} in photoionization
	calculations. To represent the outgoing Coulomb function with
	complex Gaussians, they employed Prony’s method used earlier by
	Huzinaga~\cite{Huzinaga1965}. The main drawback of this
	method is that it cannot cover all the necessary radial ranges
	because the radial grid must satisfy a specific square root
	distribution. This is why in~\cite{Matsuzaki2} a clear
	difference is visible between the exact continuum function and its
	complex fitting in the range $0<r<5~a.u.$. Matsuzaki \textit{et al}
	also optimized complex Gaussian sets to fit  complex Slater
	functions~\cite{Matsuzaki3}. The first aim of the
	present study is to explore further the capacity of complex
	Gaussians sets to represent continuum states by proposing and
	implementing a full nonlinear optimization method. In order to
	test its limitations and range of applicability we take, as an
	illustration, a set of regular radial Coulomb functions with
	different energies; the same can be done with any one-electron
	continuum function as, for example, a numerically generated
	distorted wave. We also highlight situations where real Gaussians
	become insufficiently accurate, and point out the advantage of
	using complex Gaussians instead.
	
	The second purpose of this manuscript is to demonstrate
	the advantage of such Gaussian representation when evaluating, for
	example, ionization cross sections. The key idea is to propose an
	all-Gaussian approach which allows one to analytically evaluate
	all necessary integrals. It is very well
	known that using Gaussians to represent bound states renders most
	of the necessary integrals analytical, a feature that is
	especially important in the molecular multicenter case by
	applying the Gaussian product theorem. One question naturally
	arises: is it possible to exploit such analytical advantages also
	when continuum states are involved? If they are represented by
	standard real Gaussians, the answer is obviously positive. What
	about complex Gaussians? Is the Gaussian product theorem still
	applicable? It can be easily verified that when complex exponents
	are present, the product of two Gaussians with different centers
	will lead to a Gaussian centered at some complex position. Clearly
	the latter has no physical sense, but it is not a mathematical
	obstacle for performing multicenter integrals as shown by Kuang
	and Lin~\cite{kuang1997,kuang1997b}. While our
	long-term goal is to deal with molecular systems, as a
	first step we consider here the one-center, atomic, case to put
	our all-Gaussian proposal on solid grounds. To this
	effect we consider some one-electron matrix elements
	$\braket{\psi_{\mathbf{k_e}}^-(\mathbf{r})|\widehat{O}|\phi_{i}(\mathbf{r})}$
	corresponding to a transition from an initial state
	$\phi_{i}(\mathbf{r})$ to a continuum state
	$\psi_{\mathbf{k_e}}^-(\mathbf{r})$ with ejected electron's
	momentum $k_e$; the bound-continuum transition occurs \emph{via}
	an ionization process represented by some operator $\widehat{O}$.
	To envisage an all-Gaussian integration approach, the main
	difficulty stands in a numerically robust representation
	of the continuum wavefunction by a set of Gaussian functions, and
	this up to a sufficiently large radial distance. We shall show that
	thereafter all integrations can be
	performed analytically with either real or complex sets. As
	indicated above, we consider here the one-center case and take
	hydrogen as a benchmark since its ionization cross sections 
	are known exactly for the photon impact case and also, within the first
	Born approximation, for the electron impact case.
	
	In section~\ref{sec:level2} we present the algorithm that performs
	the fitting. The numerical method, based on the approach of
	Nestmann and Peyerimhoff~\cite{nestmann},  has been
	improved here on several aspects, by using a better optimization
	method and extending it to deal with complex Gaussians. We compare
	different fitting options with real or complex Gaussians and we
	point out the advantage of the latter. In section~\ref{sec:level3}
	we illustrate the approach in two benchmark applications, the
	ionization of hydrogen by impact of either an electron or a
	photon. In both cases we compare the exact cross sections with
	those calculated with real or complex Gaussian fits of Coulomb
	continuum states. A brief conclusion is presented in
	section~\ref{sec:level6}. Atomic units are used unless indicated
	otherwise.
	
	
	\section{ \sffamily \Large Fitting with complex Gaussians \label{sec:level2}}
	
	We wish to develop an efficient approach that fits complex Gaussians to represent a
	set of continuum functions arising, for example, in ionization calculations. We
	start with a detailed comparison between using real and complex Gaussians to
	highlight the potential benefits of the latter.
	
	
	\subsection{\sffamily \Large \label{subsec1:level2}Fitting strategy}
	
	We aim to approximate a set of arbitrary functions $f_{\eta}(r)$, $\eta=1,\dots,\eta_{max}$
	by a linear combination of
	$N$ Gaussians:
	\begin{equation}
	f_{\eta}(r) \approx f_{\eta}^G(r) = \sum_{i=1}^N [c_i]_{\eta} \exp(-\alpha_i r^2) \text{.}
	\end{equation}
	To do so, in the case of real Gaussians, Nestmann and Peyerimhoff~\cite{nestmann}
	proposed a least square approach which consists in minimizing, on some radial
	grid $\{r_{\kappa}\}_{{\kappa}=1,\dots,{\kappa}_{max}}$, the function:
	\begin{align}
	\Xi(\alpha_1,\dots,\alpha_N) =
	\sum_{\eta} \frac{\sum_{\kappa} \left( f_{\eta}(r_{\kappa}) - f_{\eta}^G(r_{\kappa}) \right)^2}
	{\sum_{\kappa} \left( f_{\eta}(r_{\kappa}) \right)^2}
	+ D(\alpha_1,\dots,\alpha_N) \text{.}
	\label{FlRNC0}
	\end{align}
	The $\Xi$ function depends on $N$ nonlinear parameters, the
	exponents $\{\alpha_i\}_{i=1,\dots,N}$ and $\eta_{max} \times N$ linear parameters, the
	expansion coefficients $\{[c_i]_{\eta}\}_{i=1,\dots,N,\eta=1,\dots,
		\eta_{max}}$. In~\cite{nestmann} the standard Powell method~\cite{powell} is
	used to optimize
	the exponents, while the linear coefficients are optimized by a standard least square
	method. Iterations are performed to alternate those two optimizations: after each variation of
	the exponents the coefficients are updated using least squares, and the process is repeated
	until convergence to a local minimum is reached. In eq.~(\ref{FlRNC0}) a penalty function is
	added to avoid the convergence of two exponents to the same value. It is defined as:
	\begin{equation}
	D(\alpha_1,\dots,\alpha_N) = \sum_{i=2}^{N} \sum_{j=1}^{i-1}
	\exp \left( -g \left| \frac{\alpha_i}{\alpha_j}
	- \frac{\alpha_j}{\alpha_i} \right| \right) \text{,}
	\label{Daddedtofit0}
	\end{equation}
	where $g$ is a fixed parameter (generally $g \approx r_{\kappa_{max}}$).
	
	Here, we generalize the approach of Nestmann and Peyerimhoff~\cite{nestmann} for
	complex exponents $\alpha_i = \Re(\alpha_i) + i \Im(\alpha_i) $, with
	$\Re(\alpha_i)>0$. The optimization function $\Xi$ becomes:
	\begin{equation}
	\Xi(\Re(\alpha_1),\dots,\Re(\alpha_N),
	\Im(\alpha_1),\dots,\Im(\alpha_N)) =
	\sum_{\eta} \frac{\sum_{\kappa} | f_{\eta}(r_{\kappa}) - f_{\eta}^G(r_{\kappa}) |^2}
	{\sum_{\kappa} | f_{\eta}(r_{\kappa}) |^2}
	+ D(\Re(\alpha_1),\dots,\Re(\alpha_N)) \text{,}
	\label{FlRNC}
	\end{equation}
	and now depends on $2N$ non-linear real
	parameters $\{\Re(\alpha_i),\Im(\alpha_i) \}_{i=1,\dots,N}$ so that $\Xi$ is seen as a map
	from $\mathbb{R}^{2N}$ to $\mathbb{R}$. The penalty function is the same
	as in eq.~(\ref{Daddedtofit0}), applied only to the real part of the
	exponents. eq.~(\ref{FlRNC0}) is a particular case of eq.~(\ref{FlRNC}) when
	the exponents $\{\alpha_i\}$ and the coefficients $\{c_i\}$ are real. In order
	to minimize the fitting error $\Xi$, we choose to optimize the exponents $\{\alpha_i\}$ by
	using the Bound Optimization BY Quadratic Approximation (BOBYQA)~\cite{BOBYQA}, still
	alternating with a least square optimization of the coefficients $\{c_i\}$. Both Powell
	and BOBYQA are gradient free methods and attempt to find a local minimum. Since $\Xi$ has
	many local minima, the aim of the numerical optimization is to find a local minimum that gives
	a reasonable fitting accuracy. A critical issue in both methods is the choice of the initial
	values of the exponents. For BOBYQA, in addition to this, we have to fix the
	initial ($\Delta_i$) and final ($\Delta_f$) trust region radii where $\Xi$ is
	approximated to a quadratic model. The optimization is stopped when the Euclidean
	dimension of the step is less or equal to $\Delta_f$. On the other hand, the optimization
	with Powell is stopped when no further improvement is obtained after varying the
	exponents. A supplementary condition to stop the optimization may be the value of $\Xi$ or
	the CPU time. The main difference between these two methods is that in the case of the standard Powell algorithm, we first determine the search
	directions and then find the optimal step along those directions,
	whereas using BOBYQA, we first set the step (by choosing the
	trust region) and then the directions are found in order to
	improve the quadratic model or minimize the objective function
	$\Xi$. For more details about the algorithms we refer the reader
	to Refs.~\cite{powell} and~\cite{BOBYQA}.

	As an illustration, we consider a set of 6 regular Coulomb
	functions~$\mathscr{E}:\{ F_{1}(r)/k_1,\dots,F_{6}(r)/k_6 \}$, defined as~\cite{MathFunc}:
	\begin{equation}
	F_{\eta}(r) = F_{l,k_{\eta}}(r) = (2k_{\eta}r)^{l+1} e^{\frac{\pi
			z}{2k_{\eta}}} \frac{\left|\Gamma\left( l+1-\frac{iz}{k_{\eta}}
		\right)\right|}{2\Gamma\left( 2l+2 \right)}
	e^{ik_{\eta}r}  \mathstrut_1 F_1 \left( l+1-\frac{iz}{k_{\eta}} , 2l+2 ; -2ik_{\eta}r \right)
	\text{,}
	\label{regularCoulombFunction}
	\end{equation}
	with wavenumbers $k_{\eta}=0.5+0.25(\eta-1)~a.u.$, and angular
	momentum number $l=1$. $\mathstrut_1 F_1$ is the Kummer confluent
	hypergeometric function. The real valued functions
	$\{F_{\eta}(r)\}$ are the exact solutions of the one particle
	Schrodinger equation with Coulomb potential $-z/r$ and are
	strongly oscillating for large positive energies. For charge $z=1$
	we have the hydrogen continuum states while if we set $z=0$ we
	obtain the spherical Bessel functions. The set $\mathscr{E}$
	serves here as a test to compare real and complex Gaussian
	fittings, and will be used also in the cross section calculations
	of section~\ref{subsec:level5}. We apply the strategy presented
	above up to $r_{\kappa_{max}}=25~a.u.$ with a radial step
	$0.025~a.u.$ to fit $\mathscr{E}$ with either $N=30$ real
	Gaussians ($30$ nonlinear real parameters to reproduce the real
	functions $F_{\eta}(r)$) or $N=30$ complex Gaussians ($60$
	nonlinear real parameters to reproduce both the real functions
	$F_{\eta}(r)$ and the imaginary part which is $0$ here), and we
	set $g=27$. The number $N$ is to be chosen
	sufficiently large as to reproduce the regular Coulomb functions
	in the considered range of energy and within the fitting box
	$\mathscr{B}$. After several convergence tests on $N$ by
	inspection of the reached $\Xi_{opt}$, we found that $N=25$
	complex Gaussians could be judged as sufficient. In the physical
	application presented in section 3.2, however, integrations go up
	to 25 a.u. and in order to reduce cross section errors, we chose
	$N=30$. For the sake of
	comparison, the same $N=30$ is taken here also for the real
	Gaussian representation. Should one consider higher energies
	and/or larger radial domains, a convergence study should be
	envisaged possibly requiring  a larger $N$.
	
	It is worth emphasizing that once the optimal set of $N$
	exponents $\{\alpha_i\}$ will be found, they can be employed to
	represent with a reasonable accuracy any other function
	$F_{\eta}(r)$ within the considered energy range $k_{\eta}\in
	[0.5;1.75]$. Indeed, simply performing a linear least squares
	method will provide the corresponding optimal coefficients $\left[
	c_i \right]_{\eta}$.
	
	
	\subsection{ \sffamily \Large \label{subsec2:level2}Limitations of real Gaussians}
	
	In this subsection we focus on the use of real Gaussians. We first show the efficiency of
	the quadratic method BOBYQA~\cite{BOBYQA} and then highlight the limitations of real Gaussians
	in representing continuum functions.
	
	\subsubsection{ \sffamily \normalsize \label{sec2:subsec2:level1}Comparison between BOBYQA and Powell}
	
	We wish all the initial exponents $\alpha_i$ to increase slowly and consistently within an interval
	$\alpha_{1}=a$ and $\alpha_{N}=b$. From our numerical experience we found that the distribution
	$\frac{\alpha_{i+1}}{\alpha_{i}} = \left( \frac{\alpha_{N}}{\alpha_{1}}\right)^{\frac{1}{N-1}}$
	leads to satisfactory results and is obtained by picking up the initial exponents
	as:
	\begin{equation}
	\ln(\alpha_i) =  \frac{1}{N-1} \left[ (N-i)\ln(a) + (i-1)\ln(b) \right] \text{.}
	\label{initlog}
	\end{equation}
	The value of $a$ should be chosen small enough to reach the end of the fitting
	box: $e^{-ar_{max}^2} \sim 1$.
	
	For the optimization with BOBYQA, we set $a=10^{-6}$, $b=1$. Two research bounds are defined
	$\alpha_{min}=10^{-6}$ and $\alpha_{max}=10$. The initial trust region is $\Delta_i=0.01$ and
	the final one is $\Delta_f=10^{-6}$. For Powell optimization there are no constraints on
	$\{ \alpha_i \}$ except being strictly positive, and slightly different initialization parameters are
	selected: $a=10^{-4}$ and $b=10$. The time taken to perform the optimization of the set
	$\mathscr{E}$ with Powell
	is $\approx 26$ times that needed with BOBYQA. The final minimum value of the error
	$\left(\Xi-D\right)$ found with Powell is $0.18 \times 10^{-3}$ and $0.02 \times 10^{-3}$
	with BOBYQA. The optimal sets of real exponents obtained by BOBYQA or Powell are displayed
	in the second and third columns of Table~\ref{tab:tablealpha}.
	
	From our numerical experience, BOBYQA is faster and more efficient in the present
	context. This may not be true for all optimizations but generally BOBYQA turns out to be at
	least as efficient as Powell method or better especially for a large number of Gaussians.
	
	
	\subsubsection{ \sffamily \normalsize \label{sec2:subsec2:level2}Deviations at large distances}
	
	Looking at the fitted functions (not shown), both optimizations lead to a fitting quality which
	is very good inside the fitting box $\mathscr{B}=[0,25]$. However, the error increases very quickly
	for~$r>25$. This is due to the fact that the price to pay for a good optimization within $\mathscr{B}$
	is the presence of small exponents $\left(\alpha_i \sim 10^{-4}\right)$ and very large associated
	coefficients $\left(c_i\sim 10^{10}\right)$. The presence of such diffused Gaussians with very
	important amplitudes at large distance may not be a serious problem in the calculation of matrix
	elements; indeed, the integration over continuum functions is usually accompanied by a decreasing
	radial exponential (factor coming from the Hermitian product with bound states) and the fitting error
	beyond some physical distance will not affect the numerical calculation. This is why we choose to
	examine the effective functions
	\begin{equation}
	F_{\eta}^{eff}(\beta,r) = F_{\eta}(r) e^{-\beta r} r^2
	\label{eq:effFunc}
	\end{equation}
	instead of $F_{\eta}(r)$ itself. The exponential corresponds to a bound decreasing
	factor where $\beta$ is a positive number and $r^2$ comes from the integration volume element in
	spherical coordinates. The deviation at large distances between the fitting and the original
	function $F_{\eta}^{eff}$ is obviously very sensitive to the value of $\beta$.
	
\begin{figure}[]
		\begin{minipage}{1\linewidth}
			\vspace{-2.25 cm}
			\begin{minipage}{0.49\linewidth}
				\includegraphics[width=\linewidth]{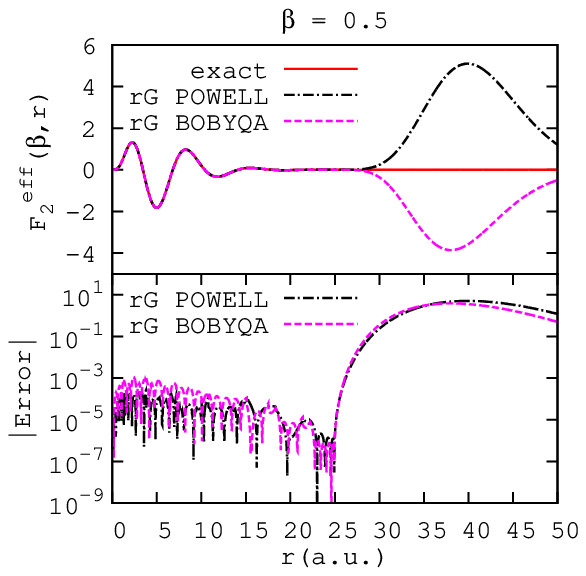}
			\end{minipage}
			\begin{minipage}{0.49\linewidth}
				\includegraphics[width=\linewidth]{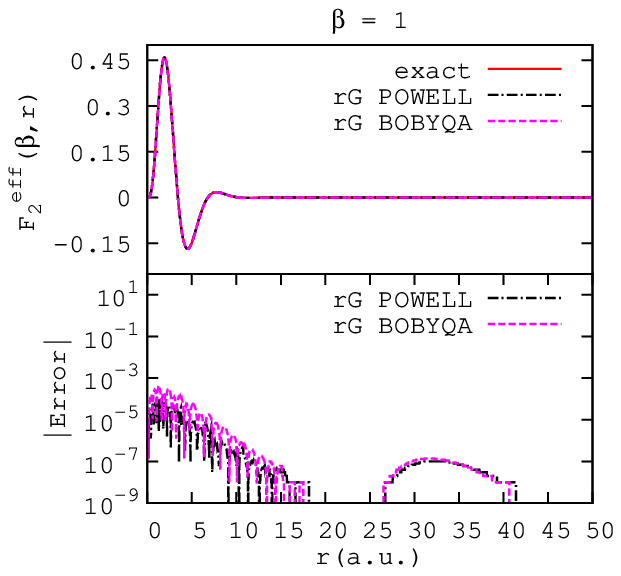}
			\end{minipage}
		\end{minipage}
		\vspace{-1.75 cm}
		\caption{\label{fig:PICSF2dec} Effective function
			$F_{2}^{eff}(\beta,r)=F_{2}(r) e^{-\beta r} r^2$ ($l=1$ and $k_2=0.75~a.u.$) and its fitting
			with real Gaussians using Powell (rG POWELL) or BOBYQA (rG BOBYQA) optimizations, for
			$\beta=0.5$ (left upper panel) and $\beta=1$ (right upper panel). The absolute
			errors are plotted in the corresponding bottom panels.}
	\end{figure}

	As an example, we examine in Figure~\ref{fig:PICSF2dec}
	the fitting of $F_{2}^{eff}(\beta,r)$ for two cases: $\beta=1$ and $\beta=0.5$, corresponding
	for example to the hydrogen $1s$ and $2s$ exponents, respectively. We clearly see that these
	fittings will cause trouble in the case $\beta=0.5$ since the deviation after $r=25$ will
	jeopardize radial integrals involving $F_2^{eff}(\beta,r)$. However when $\beta=1$ the bound
	state cancels this  deviation. For higher energies~($\eta = 3,\dots,6$) the same problem
	arises (not shown). The
	faster the functions oscillate the more important the deviation. For $\eta = 1$, on the other
	hand, the decreasing term $e^{-\beta r}$ cancels this deviation for both $\beta=0.5$ and $\beta=1$.
	
	
	\subsubsection{ \sffamily \normalsize \label{subsubsec1:subsec2:level2}Using reduced bounds (RB) for the search of the exponents}
	
\begin{figure}[]
		\vspace{-2.25 cm}
		\hspace{1.5 cm}
		\begin{minipage}{0.75\linewidth}
			\includegraphics[width=\linewidth]{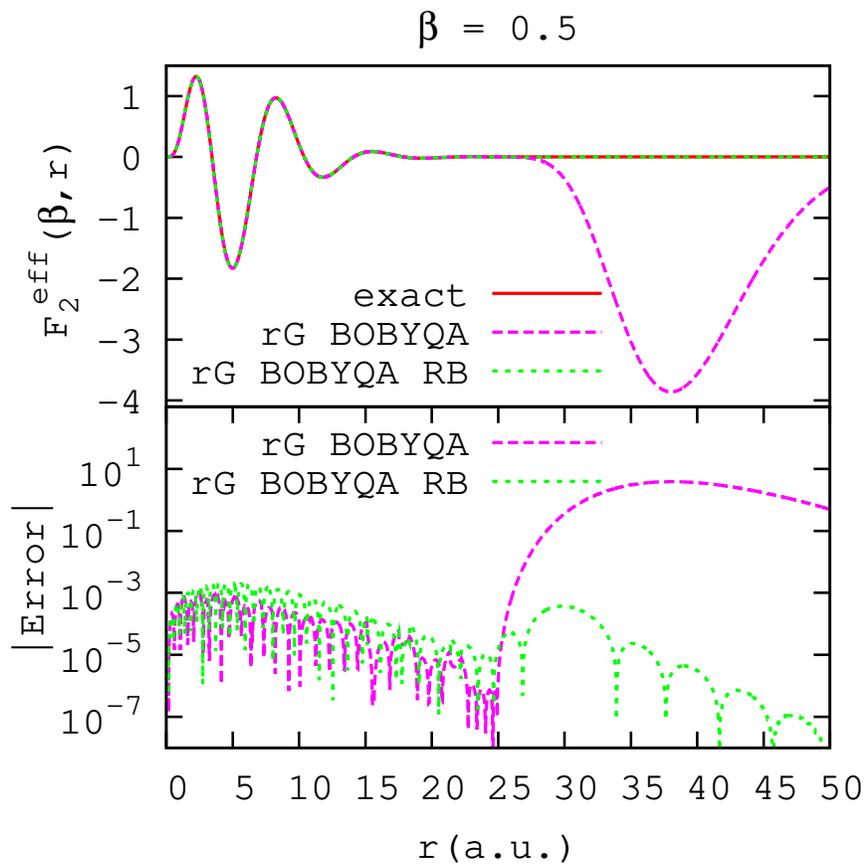}
		\end{minipage}
		\vspace{-2 cm}
		\caption{\label{fig:F34decRB}
			Upper panel: $F_{2}^{eff}=F_{2}(r) e^{-0.5 r} r^2$ with its fitting using real
			Gaussians and BOBYQA without constraint on the
			exponents (rG BOBYQA) or after reducing the bounds (rG BOBYQA RB).
			The absolute error on the fitting is plotted in the bottom panel.}
	\end{figure}

	We try in this section to soften the errors coming from the diffused part of the fitting. To
	do so with BOBYQA, the optimization lower bound is modified to $\alpha_{min}=0.01$, so that $\alpha_i$ does not fall below $0.01$. The optimal
	set of exponents in this reduced bound (RB) case is shown in the fourth column
	of Table~\ref{tab:tablealpha}. They are overall of
	the same order of magnitude as those obtained without constraint but the
	lower bound forbids the smallest, possibly troublesome, exponents.
	Figure~\ref{fig:F34decRB}
	shows the resulting improvement at large distances of $F_2^{eff}$ (RB BOBYQA) in the case
	$\beta=0.5$. While the overall fitting is better, the accuracy is slightly worse in the
	fitting box $\mathscr{B}$
	because imposing a constraint on the lower bound of $\alpha_i$ reduces the overall
	flexibility. Even in this RB approach the coefficients $\{c_i\}_{\eta}$ remain
	large. For example, the second column of Table~\ref{tab:tablecoefficients} shows the
	magnitude of the coefficients $[c_i]_{5}$ for $\eta=5$ ($k_5=1.5$). All coefficients are
	larger than~$10^6$ and most coefficients are of order~$10^{10}$ or more. When the continuum
	functions are substituted by the Gaussian combinations, this ill-conditioning generates
	a numerical error that is related to the limited machine precision.
	
	
	\subsection{ \sffamily \Large \label{subsec3:level2}Complex Gaussians}
	
	In section~\ref{subsec2:level2} we have shown that representing oscillating functions with
	real Gaussians requires very small exponents $\{\alpha_i\}$ and implies very large
	coefficients $\{c_i\}$. This causes a large deviation out of the fitting box that may turn
	out to be troublesome in a given application. A partial
	solution could be to reduce the exponents bounds but the presence of very large values
	of $\{c_i\}$ does not disappear. We will now explore the ability of complex Gaussians to
	soften these problems.
	
	In order to optimize the set $\mathscr{E}$ of functions~(\ref{regularCoulombFunction})
	defined in section~\ref{subsec1:level2}, we pick the initial complex exponents as:
	\begin{equation}
	\left\lbrace
	\begin{aligned}
	&\ln(\Re(\alpha_i)) =  \frac{1}{N-1} \left[ (N-i)\ln(a) + (i-1)\ln(b) \right] \\
	&\Im(\alpha_i) = 0
	\end{aligned}
	\right.
	\label{initcomplexexpo}
	\end{equation}
	with $a=10^{-4}$, $b=100$, and we fix the following research bounds:
	\begin{equation}
	\left\lbrace
	\begin{aligned}
	& 10^{-4} \le \Re(\alpha_i) \le 1000 \\
	& -0.1 \le \Im(\alpha_i) \le 0.1 \text{.}
	\end{aligned}
	\right.
	\label{boundcomplexexpo}
	\end{equation}
	The trust regions are $\Delta_i=0.01$ and $\Delta_f=10^{-6}$.
	
	The optimal set of complex exponents obtained for the set
	$\mathscr{E}$ are shown in the fifth column of
	Table~\ref{tab:tablealpha}. We recall that the complex
	Gaussian expansion optimizes a set of real-valued radial Coulomb
	functions. Exponents do not appear in complex conjugate pairs, thus necessarily requiring complex coefficients to build up a real function. 
	We verified that the imaginary part resulting from the complex 
	combinations of the	optimal complex Gaussians is indeed negligible.
	\begin{figure}[]
		\begin{minipage}{1\linewidth}
			\vspace{-2.50 cm}
			\begin{minipage}{0.48\linewidth}
				\includegraphics[width=\linewidth]{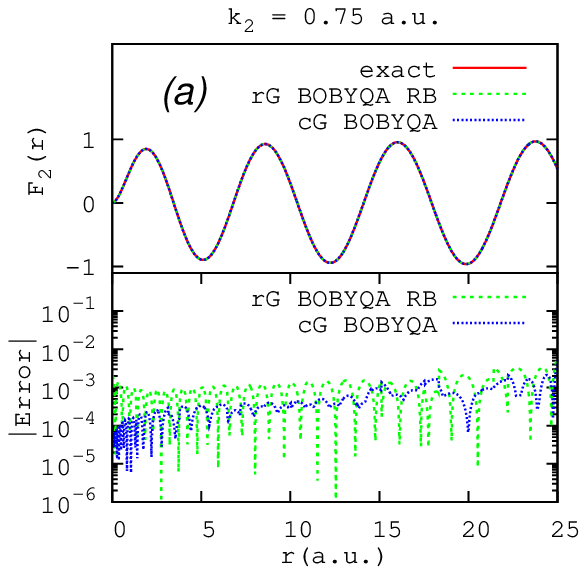}
			\end{minipage}
			\hfill
			\begin{minipage}{0.48\linewidth}
				\includegraphics[width=\linewidth]{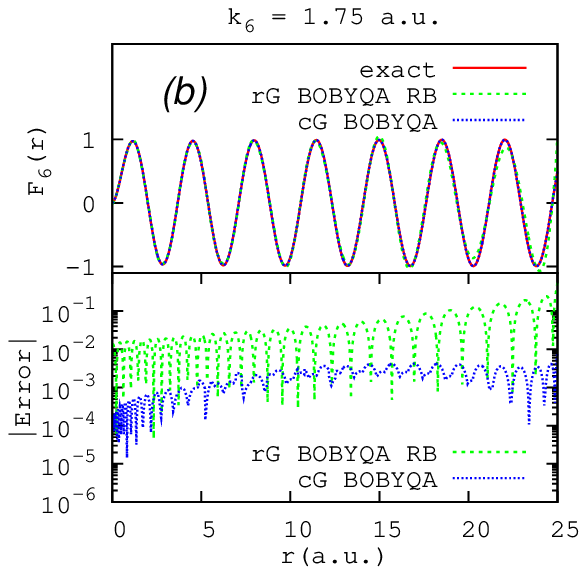}
			\end{minipage}
			\vfill
			\vspace{-1.90 cm}
			\begin{minipage}{0.48\linewidth}
				\includegraphics[width=\linewidth]{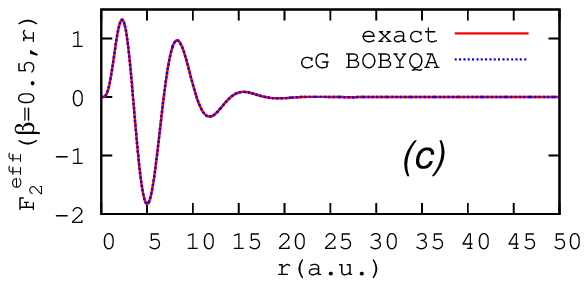}
			\end{minipage}
			\hfill
			\begin{minipage}{0.48\linewidth}
				\includegraphics[width=\linewidth]{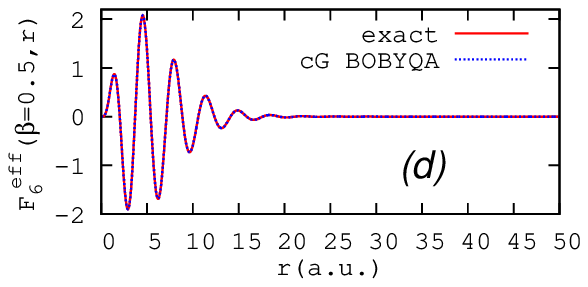}
			\end{minipage}
		\end{minipage}
		\caption{\label{fig:F236} In panels (a) and (b) the sample Coulomb function $F_{\eta}(r)$, with $l=1$,
			is plotted for $\eta=2$ and $6$ respectively with the fitting by $30$ real Gaussians using BOBYQA with
			reduced bounds (rG BOBYQA RB) or by $30$ complex Gaussians (cG BOBYQA). The associated panels show the
			corresponding absolute errors within the fitting box. In the bottom panels (c) and (d) the effective
			function $F_{\eta}^{eff}(\beta,r)$ (see eq.~(\ref{eq:effFunc})) with $\beta=0.5$ are compared to
			their cG BOBYQA representation, and this
			on a radial scale extended to $50~a.u.$.}
	\end{figure}
	Concerning the real part, Figure~\ref{fig:F236} shows two
	functions of the set, $F_{2}(r)$ and $F_{6}(r)$, with their
	fitting and the corresponding absolute errors, using: (i) $30$
	real Gaussians and BOBYQA with constraints on $\{\alpha_i\}$ (rG
	BOBYQA RB) or (ii) $30$ complex Gaussians and BOBYQA (cG BOBYQA).
	\begin{table}
		\begin{tabular}{c|c|c|c|c}
			& rG POWELL                            &  rG BOBYQA
			& rG BOBYQA RB
			&  cG BOBYQA \\
			$\alpha_1$    & $0.0000092$   & $0.0000128$
			& $0.0122973$   & $0.0001000$ $+$ $0.0151526$ $j$ \\
			$\alpha_2$    & $0.0000912$   & $0.0000244$
			& $0.0145577$   & $0.0001418$ $-$ $0.0171753$ $j$ \\
			$\alpha_3$    & $0.0009435$   & $0.0005575$
			& $0.0170202$   & $0.0013875$ $-$ $0.0021493$ $j$ \\
			$\alpha_4$    & $0.0019943$   & $0.0016484$
			& $0.0197536$   & $0.0019035$ $+$ $0.0216135$ $j$ \\
			$\alpha_5$    & $0.0027596$   & $0.0034884$
			& $0.0228103$   & $0.0025827$ $+$ $0.0298661$ $j$ \\
			$\alpha_6$    & $0.0044246$   & $0.0084723$
			& $0.0262398$   & $0.0034908$ $-$ $0.0241050$ $j$ \\
			$\alpha_7$    & $0.0057793$   & $0.0111337$
			& $0.0301176$   & $0.0046754$ $-$ $0.0338441$ $j$ \\
			$\alpha_8$    & $0.0072420$   & $0.0136592$
			& $0.0344988$   & $0.0061876$ $+$ $0.0378398$ $j$ \\
			$\alpha_9$    & $0.0089783$   & $0.0163089$
			& $0.0394645$   & $0.0156781$ $+$ $0.0504851$ $j$ \\
			$\alpha_{10}$ & $0.0114303$   & $0.0194778$
			& $0.0451376$   & $0.0196315$ $-$ $0.0395359$ $j$ \\
			$\alpha_{11}$ & $0.0141988$   & $0.0226689$
			& $0.0516205$   & $0.0246020$ $-$ $0.0352098$ $j$ \\
			$\alpha_{12}$ & $0.0178271$   & $0.0267680$
			& $0.0591117$   & $0.0308382$ $-$ $0.0302620$ $j$ \\
			$\alpha_{13}$ & $0.0221354$   & $0.0311299$
			& $0.0677411$   & $0.0386508$ $-$ $0.0436192$ $j$ \\
			$\alpha_{14}$ & $0.0272471$   & $0.0358749$
			& $0.0778144$   & $0.0484105$ $+$ $0.0624525$ $j$ \\
			$\alpha_{15}$ & $0.0341612$   & $0.0412564$
			& $0.0896070$   & $0.0618203$ $+$ $0.0513264$ $j$ \\
			$\alpha_{16}$ & $0.0436030$   & $0.0473579$
			& $0.1036178$   & $0.0816826$ $-$ $0.0552154$ $j$ \\
			$\alpha_{17}$ & $0.0622928$   & $0.0543047$
			& $0.1205385$   & $0.1936246$ $+$ $0.0097332$ $j$ \\
			$\alpha_{18}$ & $0.1058245$   & $0.0616580$
			& $0.1407606$   & $0.2915597$ $+$ $0.0024838$ $j$ \\
			$\alpha_{19}$ & $0.1361998$   & $0.0696498$
			& $0.1648189$   & $0.4971406$ $-$ $0.0113366$ $j$ \\
			$\alpha_{20}$ & $0.1729779$   & $0.0779709$
			& $0.2204008$   & $0.8532849$ $+$ $0.0200975$ $j$ \\
			$\alpha_{21}$ & $0.2157985$   & $0.0871103$
			& $0.2668776$   & $1.3797714$ $+$ $0.0062079$ $j$ \\
			$\alpha_{22}$ & $0.2755926$   & $0.0970341$
			& $0.3164855$   & $2.1958583$ $-$ $0.0311758$ $j$ \\
			$\alpha_{23}$ & $0.3476818$   & $0.1087089$
			& $0.3754007$   & $3.5672935$ $-$ $0.0087858$ $j$ \\
			$\alpha_{24}$ & $0.4425831$   & $0.1228380$
			& $0.4395371$   & $5.7419028$ $+$ $0.0012293$ $j$ \\
			$\alpha_{25}$ & $0.6059363$   & $0.1384401$
			& $0.5106744$   & $9.2535835$ $-$ $0.0212019$ $j$ \\
			$\alpha_{26}$ & $0.7983113$   & $0.1591001$
			& $0.5897495$   & $14.868739$ $-$ $0.0267870$ $j$ \\
			$\alpha_{27}$ & $1.3120641$   & $0.3614618$
			& $0.6853581$   & $23.954872$ $-$ $0.0406721$ $j$ \\
			$\alpha_{28}$ & $1.6592168$   & $0.6148974$
			& $0.7913713$   & $38.575864$ $+$ $0.0168896$ $j$ \\
			$\alpha_{29}$ & $2.0833228$   & $1.0857781$
			& $0.9151002$   & $62.088037$ $-$ $0.0059998$ $j$ \\
			$\alpha_{30}$ & $2.6339627$   & $1.5257161$
			& $1.0660043$   & $99.986651$ $+$ $0.0164035$ $j$ \\
		\end{tabular}
		\caption{
			\label{tab:tablealpha}
			Optimal exponents obtained after fitting the set $\mathscr{E}$ of Coulomb
			functions defined by eq.~(\ref{regularCoulombFunction}). rG POWELL means that the standard
			Powell method is used with real Gaussians, rG BOBYQA means that BOBYQA is used with real
			Gaussians, rG BOBYQA RB means that BOBYQA is used with reduced bounds of real
			exponents, and cG BOBYQA means that BOBYQA is used with complex Gaussians and coefficients.}
	\end{table}
	It also shows the corresponding effective functions $F_{\eta}^{eff}(\beta,r)$, with no
	deviation due to the diffused Gaussians outside the box $\mathscr{B}$. One
	can clearly see that complex Gaussians reproduce the sample Coulomb functions with a better
	accuracy, and this becomes particularly evident for large energy values. The faster the
	functions oscillate, the more gainful complex Gaussians become. All expansion coefficients
	for the $F_5^{eff}$ case, shown in the third column of Table~\ref{tab:tablecoefficients}, remain
	moderate in contrast to the real Gaussian fitting.
	\begin{table}
		\center
		\begin{tabular}{c|c|c}
			& rG BOBYQA RB  &  cG BOBYQA \\
			$c_1$    & $+0.40823965E+07$  & $-0.0767003$ $+$ $0.0491225$ $j$ \\
			$c_2$    & $-0.55724942E+08$  & $-0.0325026$ $+$ $0.1332674$ $j$ \\
			$c_3$    & $+0.33793994E+09$   & $+0.0134629$ $-$ $0.0432612$ $j$ \\
			$c_4$    & $-0.11997914E+10$   & $-0.0334812$ $+$ $0.5057190$ $j$ \\
			$c_5$    & $+0.26244301E+10$   & $+0.8455898$ $+$ $1.1476015$ $j$ \\
			$c_6$    & $-0.28184368E+10$   & $-1.4220978$ $+$ $0.5653343$ $j$ \\
			$c_7$    & $-0.29542662E+10$   & $-4.6750397$ $+$ $0.8996489$ $j$ \\
			$c_8$    & $+0.20935337E+11$   & $+3.1306684$ $-$ $2.6002539$ $j$ \\
			$c_9$    & $-0.53704332E+11$   & $-8.6221111 $ $-$ $5.9017186$ $j$ \\
			$c_{10}$ & $+0.95479441E+11$   & $-12.3193333$ $-$ $198.5520668$ $j$ \\
			$c_{11}$ & $-0.13228302E+12$   & $-246.957659$ $+$ $276.197615$ $j$ \\
			$c_{12}$ & $+0.15059598E+12$   & $+131.906416$ $+$ $98.2360837$ $j$ \\
			$c_{13}$ & $-0.14425237E+12$   & $+148.629934$ $-$ $247.936566$ $j$ \\
			$c_{14}$ & $+0.11764816E+12$   & $-14.6173162$ $+$ $76.4157364$ $j$ \\
			$c_{15}$ & $-0.82466268E+11$   & $+12.0853088$ $-$ $104.198984$ $j$ \\
			$c_{16}$ & $+0.49659844E+11$   & $+18.8413901$ $+$ $99.0639125$ $j$ \\
			$c_{17}$ & $-0.25966739E+11$   & $-17.9257132$ $+$ $1.0993553$ $j$ \\
			$c_{18}$ & $+0.11675152E+11$   & $-6.9991082$ $+$ $8.5195011$ $j$ \\
			$c_{19}$ & $-0.38808782E+10$   & $-1.8509498$ $-$ $6.3255867$ $j$ \\
			$c_{20}$ & $+0.12464857E+10$   & $+0.9407901$ $+$ $5.5765830$ $j$ \\
			$c_{21}$ & $-0.12293506E+10$   & $-1.3594724$ $-$ $5.5649285$ $j$ \\
			$c_{22}$ & $+0.10828614E+10$   & $1.3522260$ $+$ $4.9085462$ $j$ \\
			$c_{23}$ & $-0.84955303E+09$   & $-1.5426203$ $-$ $4.0800303$ $j$ \\
			$c_{24}$ & $+0.61321395E+09$   & $+1.2864865$ $+$ $3.4277871$ $j$ \\
			$c_{25}$ & $-0.36730844E+09$   & $-1.0624354$ $-$ $2.7270828$ $j$ \\
			$c_{26}$ & $+0.16987082E+09$   & $+0.7979392$ $+$ $1.9979734$ $j$ \\
			$c_{27}$ & $-0.57559060E+08$   & $-0.5300813 $ $-$ $1.2895942$ $j$ \\
			$c_{28}$ & $+0.15043766E+08$   & $+0.2829880$ $+$ $0.6875284$ $j$ \\
			$c_{29}$ & $-0.24101691E+07$   & $-0.1071319$ $-$ $0.2637591$ $j$ \\
			$c_{30}$ & $+0.17331481E+06$   & $+0.0207102$ $+$ $0.0528517$ $j$ \\
		\end{tabular}
		\caption{
			\label{tab:tablecoefficients}
			Optimal coefficients using BOBYQA to fit $F_{5}(r)$ with real Gaussians and reduced
			bounds (rG BOBYQA RB) and complex Gaussians (cG BOBYQA). Note that the number of
			digits shown here is not sufficient
			to rebuild the function in the case of (rG BOBYQA RB).}
	\end{table}
	
	
	\section{\sffamily \Large \label{sec:level3}Illustrative applications to ionization problems}
	
	For illustration purposes, we consider hereafter a one-electron description.
	Computing ionization cross sections involves the calculation of
	transition matrix elements
	$T_{i\mathbf{k_e}} =  \braket{\psi_{\mathbf{k_e}}^-(\mathbf{r})|\widehat{O}|\phi_{i}(\mathbf{r})}$
	where $\phi_{i}(\mathbf{r})$ represents the initial (bound) wavefunction and
	$\psi_{\mathbf{k_e}}^-(\mathbf{r})$ represents the final (continuum) wavefunctions of
	the ejected electron (with momentum $\mathbf{k_e}$). In order to keep the present investigation
	free of extra numerical uncertainties and easily reproducible, we choose as continuum state
	the analytical Coulomb function
	\begin{equation}
	\psi_{\mathbf{k_e}}^-(\mathbf{r}) = N(a)
	\frac{e^{i\mathbf{k_e}\mathbf{r}}}{(2\pi)^{\frac{3}{2}}}
	\mathstrut_1 F_1 \left( -ia , 1 ; -i(k_e r + \mathbf{k_e}\mathbf{r} ) \right) \text{,}
	\label{continuumwavefunction}
	\end{equation}
	where $N(a)=e^{\frac{\pi a}{2}}\Gamma(1+ia)$ with the Sommerfeld parameter $a=z/k_e$
	and $z$ the charge seen by the ejected electron. $\widehat{O}$ is the
	transition operator that connects the initial to final
	states: $\frac{4 \pi}{q^2} e^{i \mathbf{q} \cdot \mathbf{r}}$ in the case of
	particle impact ($\mathbf{q}$ is the momentum transfer vector)
	and $-\mathbf{\hat{\epsilon}} \cdot \mathbf{r}$ for photoionization in length
	gauge ($\mathbf{\hat{\epsilon}}$ is the polarization vector). In what follows, we will
	show that if the radial parts of
	both $\psi_{\mathbf{k_e}}^-$ and $\phi_{i}$ are expanded in Gaussians, the calculation of
	the transition matrix elements becomes  analytical for both processes. As mentioned in
	the introduction the ultimate goal is to implement such an all-Gaussian approach to
	treat scattering from polyatomic
	molecules. Here, in order to illustrate the feasibility and the numerical robustness, we
	consider first an atomic case with an initial wavefunction given by:
	\begin{equation}
	\phi_i(\mathbf{r}) = R_{n_il_i}(r) Y_{l_i}^{m_i}(\hat{r}) \text{,}
	\label{boundstatewavefunctio}
	\end{equation}
	where $n_i$,$l_i$,$m_i$ are the usual quantum numbers. For the numerical illustration, we
	shall take as benchmark the hydrogen atom ($z=1$) for which exact cross sections are available
	and serve as a solid benchmark.
	
	
	\subsection{\sffamily \Large  \label{subsec:level4}Hydrogen ionization by electron impact}
	
	We consider the ionization of a hydrogen atom by electron
	impact: $ \text{e}^- + \text{H} \rightarrow \text{H}^+ + 2\text{e}^- $. In the first
	Born approximation the colliding electron is described by a plane wave
	before (momentum~$\mathbf{k_i}$) and after the collision (momentum~$\mathbf{k_s}$), while
	the wavefunction of the ejected electron is the Coulomb
	function~$\psi_{\mathbf{k_e}}^-(\mathbf{r})$ of eq.~(\ref{continuumwavefunction}). The cross
	section calculation involves the transition matrix element
	\begin{equation}
	T_{i\mathbf{k_e}} = \frac{4\pi}{q^2} F_{i\mathbf{k_e}}(\mathbf{q}) \text{,}
	\label{transitionmatrixelements1}
	\end{equation}
	where $\mathbf{q}=\mathbf{k_i}-\mathbf{k_s}$ is the momentum transfer vector and
	\begin{equation}
	F_{i\mathbf{k_e}}(\mathbf{q})=
	\braket{\psi_{\mathbf{k_e}}^-|e^{i\mathbf{q}\mathbf{r}}|\phi_i}
	\label{atomicformfactor}
	\end{equation}
	is the atomic form factor with $\phi_i$ the initial wavefunction~(\ref{boundstatewavefunctio}).
	
	The standard way to separate angular and radial variables is to use a partial wave expansion
	of the whole continuum wavefunction~(\ref{continuumwavefunction}) over the spherical
	harmonics $Y_l^{m*}(\hat{r})Y_l^{m}(\hat{k_e})$:
	\begin{equation}
	\psi_{\mathbf{k_e}}^-(\mathbf{r}) = \sqrt{\frac{2}{\pi}} \sum_{l,m} i^l
	e^{i \delta_l} \frac{F_{l,k_e}(r)}{kr} Y_l^{m*}(\hat{r}) Y_l^{m}(\hat{k_e}) \text{,}
	\label{newCoulombContinuum}
	\end{equation}
	where $F_{l,k_e}(r)$ is the regular radial Coulomb
	function~(\ref{regularCoulombFunction}) and
	$ \delta_l = \text{arg} \left( \Gamma(l+1+\frac{zi}{k_e}) \right) $ is the
	Coulomb phase shift. Alternatively, one can extract from the
	wavefunction~(\ref{continuumwavefunction}) the highly oscillating behavior of
	the plane wave $e^{i \mathbf{k_e} \mathbf{r}}$  and expand in partial waves only the
	distortion factor represented by the confluent hypergeometric function. The hydrogen
	continuum state can then be written as
	\begin{equation}
	\psi_{\mathbf{k_e}}^-(\mathbf{r}) =
	\frac{e^{i\mathbf{k_e}\mathbf{r}}}{(2\pi)^{\frac{3}{2}}}
	\sum_{l,m} D_{l,k_e}(r) Y_l^{m*}(\hat{r}) Y_l^{m}(\hat{k_e}) \text{,}
	\label{psitotwithDl}
	\end{equation}
	where the radial factors~\cite{Dexpansion}
	\begin{equation}
	D_{l,k_e}(r) = 4 \pi e^{\frac{\pi}{2k_e}}
	\frac{\Gamma\left( 1+\frac{i}{k_e} \right)
		\Gamma\left( l-\frac{i}{k_e} \right) (-i)^l}{\Gamma\left(-\frac{i}{k_e} \right)
		\Gamma(l+1) (2l+1)!!} (k_er)^l \,
	\mathstrut_1 F_1 \left( l-\frac{i}{k_e} , 2l+2 ; -2 i k_e r \right)
	\label{Dlke1}
	\end{equation}
	are complex functions. Fiori and Miraglia~\cite{fiori} proposed
	to follow this second path and found that
	taking $l_{max}=8$ provides a sufficient accuracy in the calculation of hydrogen ionization
	by proton impact in a given energy range. If both $D_{l,k_e}(r)$ and the radial part
	of $\phi_i$ are represented by Gaussian combinations, then the transition matrix
	integral becomes analytical.
	
\begin{figure}
		\center
		\begin{minipage}{0.6\linewidth}
			\includegraphics[width=\linewidth]{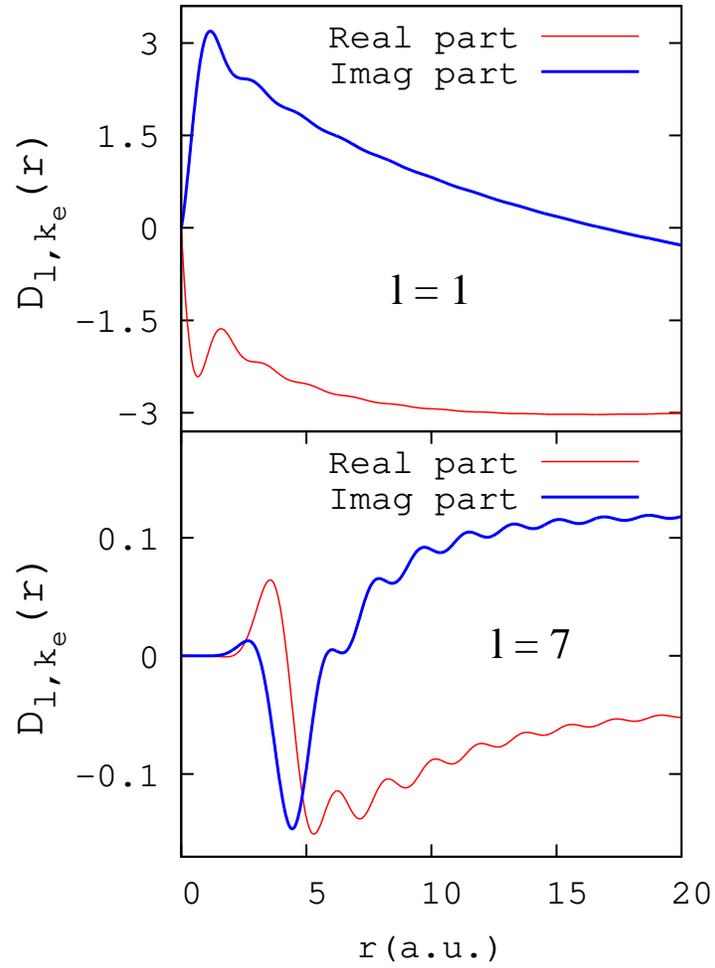}
		\end{minipage}
		\vspace{- 1.75 cm}
		\caption{\label{fig:D_l1l7k3} Real and imaginary parts of the $D_{l,k_e}(r)$ functions
			defined by eq.~(\ref{Dlke1}) for $k_e=1.75$ and $l=1$ (upper panel)
			or $l=7$ (lower panel).}
	\end{figure}

	We should first shed light on the behavior of the radial $D_{l,k_e}(r)$
	functions which, supposedly without strong oscillations~\cite{fiori}, should be easier
	to represent through a Gaussian set. Actually they are smooth only for low $l$. Functions with
	large values of $l$ manifest
	non negligible oscillations as illustrated by Figure~\ref{fig:D_l1l7k3}
	which shows the behavior of $D_{1,k_e}(r)$ and $D_{7,k_e}(r)$ for $k_e=1.75$. While the
	small $l$ case can be easily represented with Gaussians, the oscillations of the
	large $l$ case result to be very difficult to reproduce. This $l$ dependence is even
	more pronounced for larger $k_e$ values. On the other hand, it is known that the first
	partial  terms are generally those contributing the most to the continuum
	wavefunction. Consequently, the efficiency of the Gaussian fitting approach depends
	on the physical application under consideration and on the number of partial waves
	needed to achieve an overall acceptable accuracy.
	
\begin{figure}
		\begin{minipage}{1.25\linewidth}
			\vspace{-3 cm}
			\begin{minipage}{1.0\linewidth}
				\includegraphics[width=\linewidth]{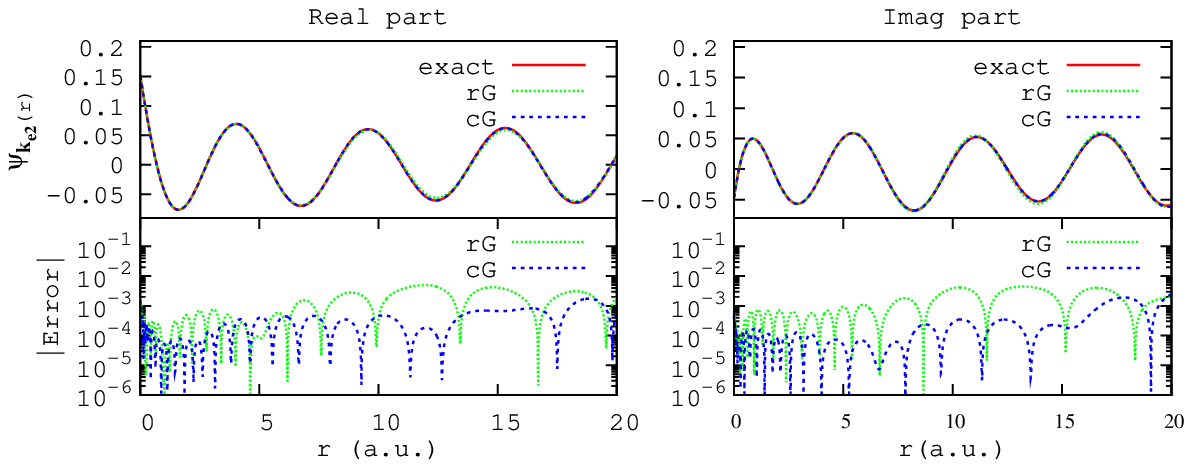}
			\end{minipage}
			\vfill
			\begin{minipage}{1.0\linewidth}
				\vspace{-4.5 cm}
				\includegraphics[width=\linewidth]{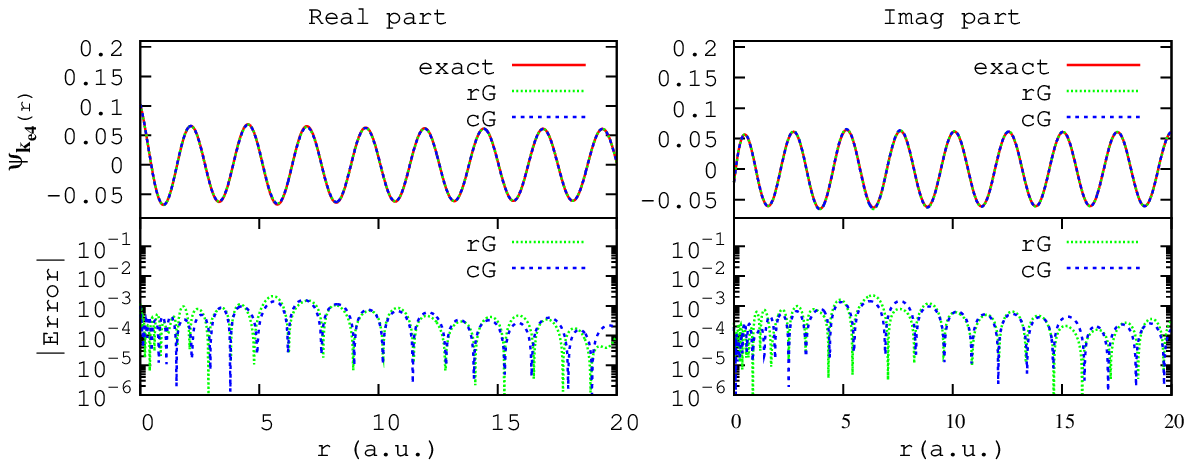}
			\end{minipage}
		\end{minipage}
		\vspace{-2 cm}
		\caption{\label{fig:psitotk2k4} Real (left panels) and imaginary (right panels)
			part of the hydrogen continuum states
			$\psi_{\mathbf{k_e}}^-(\mathbf{r})$ calculated using eq.~(\ref{psitotwithDl}) with $l$ up
			to $l_{max}=8$ and comparison with that calculated by either real Gaussians~(rG) or complex
			Gaussians~(cG), for ${k_{e}}_{2}=1.00~a.u.$ (upper panels) and ${k_{e}}_{4}=2.50~a.u.$ (bottom
			panels), and $(\hat{r},\hat{k_e})=0$ in both cases. The corresponding absolute error on
			the fitting is shown in the bottom sub-panels.}
	\end{figure}

	To compare real and complex Gaussians, we use the strategy presented in
	section~\ref{subsec1:level2} for $9$ different sets of
	distortion functions,  $\mathscr{D}_l: \{D_{l,{k_{e}}_1},D_{l,{k_{e}}_2},\dots,D_{l,{k_{e}}_5}\}$
	for  ${l=0,1,\dots,8}$. In each set we consider $5$ functions corresponding
	to ${k_{e}}_{\eta}=0.25+0.75(\eta-1)$. The fitting is performed up to $r=20$ and we set the
	parameter $g=18$ in eq.~(\ref{Daddedtofit0}). Concerning the fitting with real
	Gaussians, for a given $l$ we optimize
	$\{\Re\left(D_{l,{k_{e}}_1}\right),
	\Re\left(D_{l,{k_{e}}_2}\right),\dots, \Re\left(D_{l,{k_{e}}_5}\right)
	\}$ using $20$ real Gaussians and
	$\{\Im\left(D_{l,{k_{e}}_1}\right),
	\Im\left(D_{l,{k_{e}}_2}\right),\dots, \Im\left(D_{l,{k_{e}}_5}\right)\}$
	with $20$ other real Gaussians. On the other hand, in the complex case
	we use $20$ complex Gaussians to fit the set $\mathscr{D}_l$. In order to somehow take into account the $r^l$ behavior of $D_{l,k_e}(r)$ close to $r=0$, we add 	a $r^{\gamma_l}$ factor in the Gaussian expansion,
	\begin{equation}
	D_{l,k_e}(r) = r^{\gamma_l}
	\sum_s \left[ c_s \right]_{l,k_e} \exp\left( - \left[ \alpha_s \right]_{l} r^2  \right)
	\label{DlwithGaussians}
	\end{equation}
	where $\gamma_0=0$ and $\gamma_l=1$ for $l>0$. We use ${\gamma_l} = 1$ instead of ${\gamma_l} = l$ as a compromise that gives the vanishing behavior close to $r=0$ for $l> 0$ while avoiding error amplifications at large radial distances for large values of $l$. Figure~\ref{fig:psitotk2k4}
	shows the hydrogen continuum state at two energies ${k_{e}}_{2}=1$
	and ${k_{e}}_{4}=2.5$, calculated using eq.~(\ref{psitotwithDl}) up to $l_{max}=8$, for an
	angle $(\hat{r},\hat{k_e})=0$, and with $D_{l,k_e}(r)$ represented by either real or complex
	Gaussians using eq.~(\ref{DlwithGaussians}). The corresponding error on the fitting is
	also shown. For small energies the errors corresponding to the complex fitting are smaller. For
	large energies this advantage is lost since the partial waves with high order $l \gtrsim 3$
	can not be reproduced accurately either by real or by complex Gaussians. Nevertheless the
	continuum wavefunction is overall well reproduced with both options because the first few
	terms $-$ which contribute the most to $\psi_{\mathbf{k_e}}^-$ $-$ are
	sufficiently well fitted. In summary, with expansion~(\ref{psitotwithDl}) the fitting
	difficulties due to oscillations are not completely removed as suggested
	in~\cite{fiori} and we should remain aware of this weakness depending on the
	application. In the numerical illustration presented hereafter, the difficulty appears
	only in partial wave terms that do not contribute substantially and therefore do not
	affect the overall cross section calculation.
	
\begin{figure}
		\begin{minipage}{1.0\linewidth}
			\begin{minipage}{0.33\linewidth}
				\includegraphics[width=\linewidth]{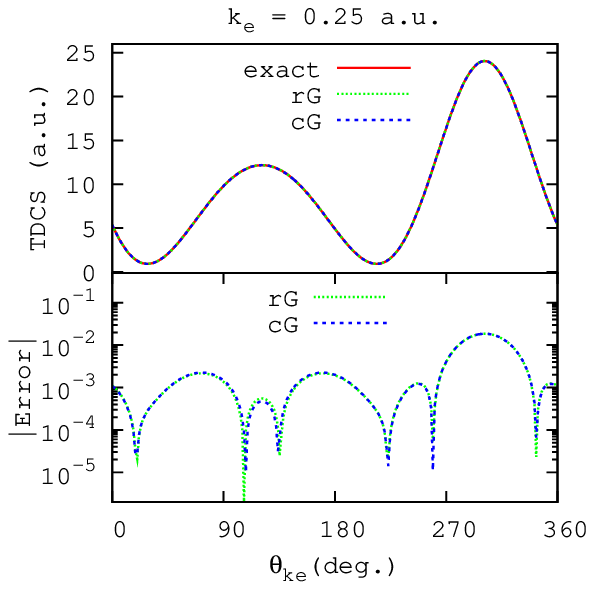}
			\end{minipage}
			\hspace{-0.5 cm}
			\hfill
			\begin{minipage}{0.33\linewidth}
				\includegraphics[width=\linewidth]{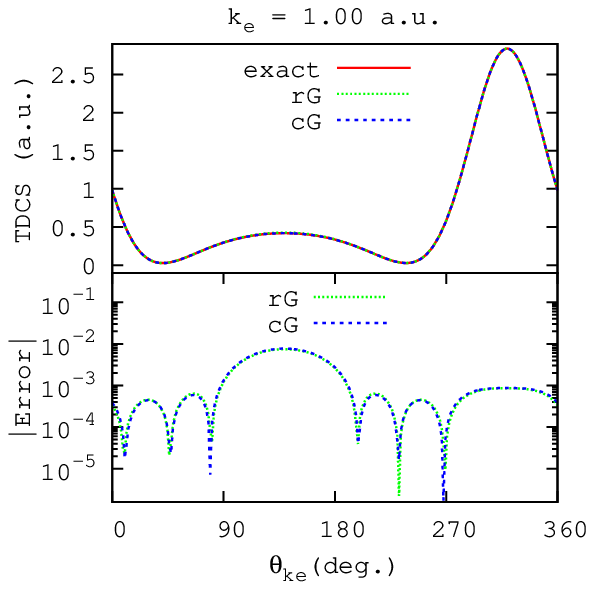}
			\end{minipage}
			\hspace{-0.5 cm}
			\hfill
			\begin{minipage}{0.33\linewidth}
				\includegraphics[width=\linewidth]{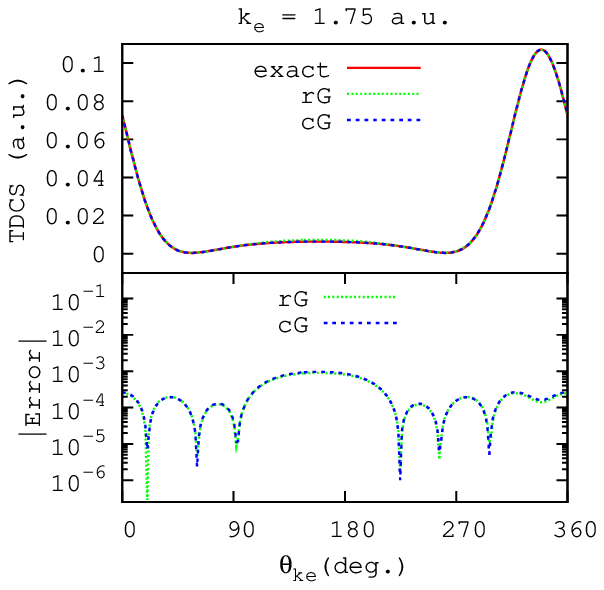}
			\end{minipage}
		\end{minipage}
		\vspace{-1.25 cm}
		\caption{\label{fig:TDCS} Upper panels: Triple differential cross section (TDCS) for
			the ionization of a hydrogen atom in the ground state by a $250$ eV electron, in
			coplanar geometry, with a scattering
			angle of $3^{\circ}$ and for ejected electron momentum $k_e=0.25$, $1.00$, or $1.75~a.u.$: Solid
			curves
			represent the exact analytical TDCS computed with eq.~(\ref{TDCSanalyt}), and dotted/dashed
			curves represent the TDCS calculated with either real Gaussians (rG) or complex
			Gaussians (cG). The corresponding absolute error is shown in the bottom panels.}
	\end{figure}

	Let us now turn to the calculation of the atomic form factor~(\ref{atomicformfactor})
	employing Gaussian expansion~(\ref{DlwithGaussians}). The bound radial function
	of eq.~(\ref{boundstatewavefunctio}) is also represented with either real
	or complex Gaussians:
	\begin{equation}
	R_{n_il_i}(r) = \sum_{t} b_t e^{-\beta_t r^2} \text{.}
	\label{boundstateGaussians}
	\end{equation}
	Using expansions~(\ref{psitotwithDl}),~(\ref{DlwithGaussians}) and~(\ref{boundstateGaussians}), the
	form factor becomes:
	\begin{equation}
	F_{i\mathbf{k_e}}(\mathbf{q}) = \frac{1}{(2\pi)^{\frac{3}{2}}}
	\sum_{l,m} Y_{l}^{m *}(\hat{k_e})
	\sum_{s,t} \left[ c_s \right]_{l,k_e}^* b_t
	\int d\mathbf{r} \, r^{\gamma_l}  \, e^{i \mathbf{Q} \mathbf{r}}
	e^{- \left( \left[ \alpha_s \right]_{l}^* + \beta_t \right)  r^2 }
	Y_{l_i}^{m_i}(\hat{r}) Y_{l}^{m}(\hat{r}) \text{,}
	\label{atomicfactorbeforeGaussians}
	\end{equation}
	where $\mathbf{Q}=\mathbf{q}-\mathbf{k_e}$ is the momentum of the hydrogen ion after the
	collision. To separate angular and radial variables we use the Rayleigh expansion:
	\begin{equation}
	e^{i \mathbf{Q} \mathbf{r}} = 4 \pi \sum_{\lambda,\mu} i^{\lambda} j_{\lambda}(Qr)
	Y_{\lambda}^{\mu*}(\hat{Q}) Y_{\lambda}^{\mu}(\hat{r}) \text{,}
	\end{equation}
	where $j_{\lambda}$ are the spherical Bessel
	functions. Therefore, in eq.~(\ref{atomicfactorbeforeGaussians}) we
	have an integral over  $3$ spherical harmonics multiplied by the following radial
	integral,
	\begin{equation}
	\mathcal{I}^{rad} = \int_0^{\infty} dr \, r^{2+\gamma_l} \,
	e^{- \left( \left[ \alpha_s \right]_{l}^* + \beta_t \right)  r^2 } \,
	j_{\lambda}(Qr)
	\end{equation}
	which can be calculated analytically (eq.~$6.6.31$ of~\cite{gradshteyn2007}):
	\begin{equation}
	\mathcal{I}^{rad} = \frac{\frac{\sqrt{\pi}}{4} \left( \frac{Q}{2} \right)^{\lambda}
		\Gamma \left( \frac{\lambda+\gamma_l+3}{2} \right)}
	{\Gamma \left( \lambda + \frac{3}{2} \right)}
	\left( \left[ \alpha_s \right]_{l}^* + \beta_t \right)^{-\frac{\lambda+\gamma_l+3}{2} }
	\,
	\mathstrut_1 F_1 \left( \frac{\lambda+\gamma_l+3}{2} , \lambda + \frac{3}{2} ;
	\frac{-Q^2}{4\left( \left[ \alpha_s \right]_{l}^* + \beta_t \right)} \right) \text{.}
	\end{equation}
	Finally the form factor~(\ref{atomicfactorbeforeGaussians}) can be written as
	\begin{equation}
	F_{i\mathbf{k_e}}(\mathbf{q}) = \sqrt{\frac{2l_i+1}{32\pi}} \sum_l \sqrt{2l+1}
	\sum_{\lambda=|l-l_i|}^{l+l_i}
	\frac{\left( \frac{iQ}{2} \right)^{\lambda}
		\Gamma \left( \frac{\lambda+\gamma_l+3}{2} \right)}
	{\Gamma \left( \lambda + \frac{3}{2} \right)}
	\sqrt{2\lambda+1} \mathscr{S}_{l,\lambda}^{rad} \mathscr{S}_{l,\lambda}^{ang} \text{,}
	\label{eq:Form_factor_1}
	\end{equation}
	where
	\begin{equation}
	\mathscr{S}_{l,\lambda}^{rad} = \sum_{s,t}
	\left[ c_s \right]_{l,k_e}^* b_t
	\left( \left[ \alpha_s \right]_{l}^* + \beta_t \right)^{-\frac{\lambda+\gamma_l+3}{2}}
	\mathstrut_1 F_1 \left( \frac{\lambda+\gamma_l+3}{2} , \lambda + \frac{3}{2} ;
	\frac{-Q^2}{4\left( \left[ \alpha_s \right]_{l}^* + \beta_t \right)} \right)
	\label{eq:Form_factor_2}
	\end{equation}
	and
	\begin{equation}
	\mathscr{S}_{l,\lambda}^{ang} = \tj{l}{l_i}{\lambda}{0}{0}{0}
	\sum_{m=-l}^l \tj{l}{l_i}{\lambda}{m}{m_i}{-m-m_i}
	Y_l^{m*}(\hat{k_e}) Y_{\lambda}^{-(m+m_i)*}(\hat{Q})
	\label{eq:Form_factor_3}
	\end{equation}
	where $\tj{j_1}{j_2}{j_3}{m_1}{m_2}{m_3}$ denote the $3j$ Wigner coefficients. Assuming that
	the hydrogen target is in its ground state $1s$, it is possible to compare to the exact
	atomic factor~\cite{Dowell}:
	\begin{align}
	F_{1s \mathbf{k_e}}(\mathbf{q}) = \frac{2\sqrt{2} e^{\frac{\pi}{2k_e}}
		\Gamma \left( 1-\frac{i}{k_e} \right)  }
	{\pi (1+Q^2)^2 \mathsf{U}^{\frac{i}{k_e}}}
	\left[ \left( 1-\frac{i}{k_e} \right) + \frac{k_e+i}{k_e \mathsf{U}}  \right] \text{,}
	\label{TDCSanalyt}
	\end{align}
	where
	\begin{equation}
	\mathsf{U} = \frac{q^2-(k_e+i)^2}{1+Q^2} \text{.}
	\label{eq:U}
	\end{equation}
	The triple differential cross section (TDCS) is defined as:
	\begin{equation}
	\frac{d^3\sigma}{d\Omega_s d\Omega_e dE_e} =
	\frac{1}{4 \pi^2}  \frac{k_s k_e}{k_i}
	|T_{1s \mathbf{k_e}}(\mathbf{q})|^2 \text{,}
	\label{eq:TDCS}
	\end{equation}
	with $E_e=k_e^2/2$ the energy of the ejected electron. $\Omega_e$ and $\Omega_s$ are the
	solid angles for the ejected and the scattered electron, respectively. We have calculated
	the TDCS for coplanar geometry at $E_i=k_i^2/2=250$ eV and scattering
	angle of $3^{\circ}$ for $k_e=0.25$, $1.00$ and $1.75$, with either real or complex
	Gaussian expansions. Both results, shown in Figure~\ref{fig:TDCS},
	are very close to the exact values given by eq.~(\ref{TDCSanalyt}), and display a similar
	accuracy. As discussed before, this can be related to the fact
	that the first partial functions $D_{l,k_e}(r)$ (making the largest contribution to the
	TDCS) are relatively smooth and can be well reproduced by both real and complex Gaussians
	with an equal precision. The very good reproduction of the exact cross section is also due to the
	presence of a fast decaying initial state ($e^{-r}$) so that any error from outside the fitting
	box will not affect the correct evaluation of the matrix element (see discussion in
	section~\ref{sec:level2}).
	
\begin{figure}
			\includegraphics[scale=0.75]{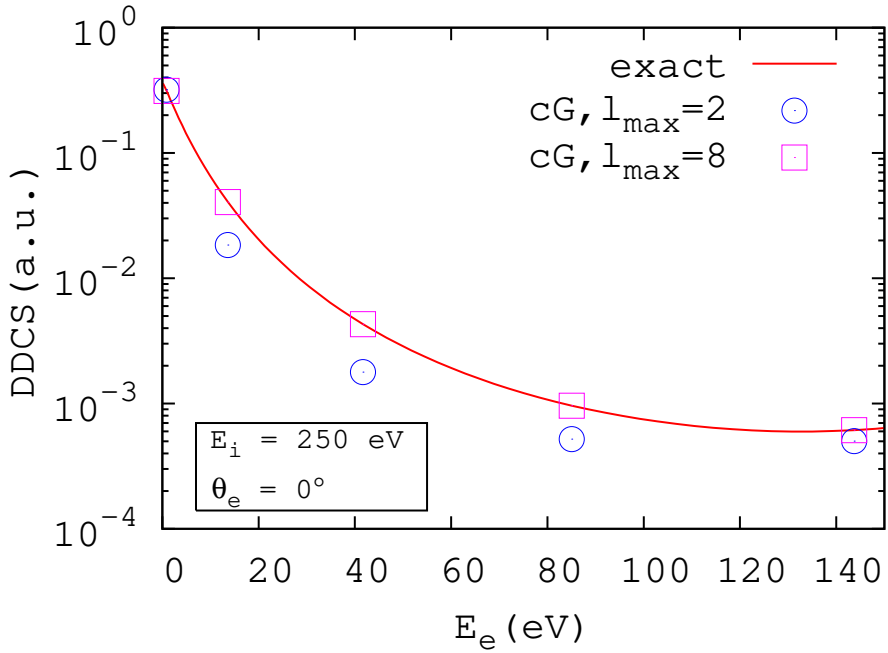}
			\caption{The doubly differential cross section (DDCS) with respect
				to the energy of the ejected electron, for an incident energy
				$E_i=250$~eV, a coplanar geometry and  ejected angle $\theta_e =
				0$. The ``exact" result (solid line) is obtained by integrating the
				TDCS~(\ref{eq:TDCS}) expressed using the closed form (Eq.~(\ref{TDCSanalyt}) and~(\ref{eq:U})),
				while the cG results (symbols) are obtained by using the partial
				wave expansion
				(Eq.~(\ref{eq:Form_factor_1}),~(\ref{eq:Form_factor_2}) and~(\ref{eq:Form_factor_3}))
				and the complex Gaussian
				representation with $l_{max}=2$ or 8.} \label{fig:DDCS}
		\end{figure}

	Other scattering quantities can be tested. For
	example, by integrating over the solid angle of the scattered
	electron, one can define a doubly differential cross section
	(DDCS)
		\begin{equation}
		\frac{d^2 \sigma}{dE_e d\Omega_{e}} = \int  \frac{d^3
			\sigma}{d\Omega_{s} d\Omega_{e} dE_e} \, d\Omega_{s} .
		\end{equation}
		The integration of the TDCS~(\ref{eq:TDCS}) is to be
		performed numerically from either the closed form of the form
		factor (Eq.~(\ref{TDCSanalyt}) and~(\ref{eq:U})) or its partial expansion
		(Eq.~(\ref{eq:Form_factor_1}),~(\ref{eq:Form_factor_2}) and~(\ref{eq:Form_factor_3}))
		up to $l_{max}$ terms that make use of the Gaussian
		representation. As an illustration, in 
		figure~(\ref{fig:DDCS}) we show the results as
		a function of the ejected energy $E_e=k_e^2/2$, for an incident
		energy $E_i=250$~eV, a coplanar geometry and  the ejected angle
		fixed at $\theta_e = 0$. We clearly observe that with $l_{max}=8$
		in the continuum state representation, the expected cross section
		is recovered (with an error of less than 1 percent).
	
	\subsection{\sffamily \Large  \label{subsec:level5}Photoionization}
	
	We now consider a hydrogen atom in the initial state $\phi_i$ illuminated by a photon
	of energy $E_{\gamma}=k_e^2/2 + V_{\text{ion}}$, where $V_{\text{ion}}$ is the
	energy needed to ionize the target. This photon interacts with the atom leading to an
	ion and a  photo-electron of energy $E_{k_e}=k_e^2/2$. The photoionization cross section
	is defined as~\cite{Burke}
	\begin{equation}
	\sigma = \frac{4 \pi^2 k_e E_{\gamma}}{c}
	\frac{1}{2l_i+1}
	\sum_{m_i}
	\int d \Omega_{\mathbf{k_e}}  \,
	\left | T_{i \mathbf{k_e}} \right |^2 \text{,}
	\label{CrossSectionPI}
	\end{equation}
	where $c$ is the speed of light in vacuum and $\Omega_{\mathbf{k_e}}$ is the solid angle of
	the ejected photo-electron. In the length gauge and the dipole approximation the transition
	matrix element is given by
	\begin{equation}
	T_{i \mathbf{k_e}} =
	\langle \psi_{\mathbf{k_e}}^-(\mathbf{r}) | -\hat{\epsilon} \cdot \mathbf{r}
	| \phi_{i}(\mathbf{r}) \rangle \text{,}
	\label{transitionmatrixelement0}
	\end{equation}
	where $\hat{\epsilon}$ defines the polarization direction. Here we expand the hydrogen
	continuum wavefunction~(\ref{continuumwavefunction}) in the standard
	way; expansion~(\ref{newCoulombContinuum}) presents the advantage that
	selection rules reduce the summation over $l$ to just $l=l_i-1$ and $l=l_i+1$. In contrast, when
	using expansion~(\ref{psitotwithDl}), converged photoionization calculations require many partial
	terms in order to cover large physical distances. For example we found that for
	a $1s$ hydrogen target, about $10$ partial terms $D_{l,k_e}(r)$ are needed to
	achieve a relative error of the order $0.01$, and one has to consider $\approx 30$ partial
	terms to reach a similar accuracy if the hydrogen is initially in the
	radially more extended $2s$ state. With
	expansion~(\ref{newCoulombContinuum}), only two partial terms appear in both $1s$ and $2s$ cases.
	
	The calculation of transition matrix elements~(\ref{transitionmatrixelement0}) involves
	two integrals:
	\begin{equation}
	\mathscr{J}^{ang}
	= \int  \ Y_{l}^{m*}(\hat{r}) Y_{1}^{0}(\hat{r}) Y_{l_i}^{m_i}(\hat{r}) d\Omega_{\mathbf{r}} \text{,}
	\label{angularintegrationPI}
	\end{equation}
	and
	\begin{equation}
	\mathscr{J}_l^{rad} = \int_0^{\infty}  \, F_{l,k_e}(r)
	\, R_{n_i l_i}(r) \, r^2  \, dr \text{.}
	\end{equation}
	The angular integral $\mathscr{J}^{ang}$ imposes the selection rules $l=l_i \pm 1$ and
	$m=m_i$. The radial function $R_{n_i l_i}(r)$ is fitted by real Gaussians as
	in eq.~(\ref{boundstateGaussians}), and the functions $F_{l,k_e}(r)$ are written as
	\begin{equation}
	F_{l_i \pm 1,k_e}(r) =  \sum_{s}
	\left[ c_s \right]_{l_i \pm 1,k_e} \exp(-[\alpha_s]_{l_i \pm 1} r^2 ) \text{.}
	\end{equation}
	with the different Gaussian parameters optimized to fit the
	$\mathscr{E}$ set, given in Table~\ref{tab:tablealpha}.
	Using these Gaussian representations, the radial integral becomes
	\begin{equation}
	\begin{aligned}
	\mathscr{J}_{l_i \pm 1}^{rad}
	&= \frac{\sqrt{\pi}}{4} \sum_{s,t}
	\left[ c_s \right]_{l_i \pm 1,k_e}  b_t {
		\left( [\alpha_s]_{l_i \pm 1} + \beta_t \right)}^{-\frac{3}{2}} \text{.}
	\end{aligned}
	\label{radialIntegralCSnewGaussian2}
	\end{equation}
	After summation over the magnetic numbers $m_i$, the cross section can be simply expressed as:
	\begin{equation}
	\sigma = \frac{8\pi \,E_{\gamma}}{3 \, (2l_i+1) \, k_e \, c}
	\Bigg[ l_i \left( \mathscr{J}_{l_i - 1}^{rad} \right)^2 + (l_i+1)
	\left( \mathscr{J}_{l_i + 1}^{rad} \right)^2 \Bigg]
	\end{equation}
	For a hydrogen target initially in a $ns$ state the cross section calculation involves
	only the $l=1$ partial function~(\ref{regularCoulombFunction}) which is expanded using the
	different  Gaussian sets optimized
	to fit the $\mathscr{E}$ set given in section~\ref{sec:level2}.
	
\begin{figure}
		\begin{minipage}{\linewidth}
			\begin{minipage}{0.49\linewidth}
				\includegraphics[width=\linewidth]{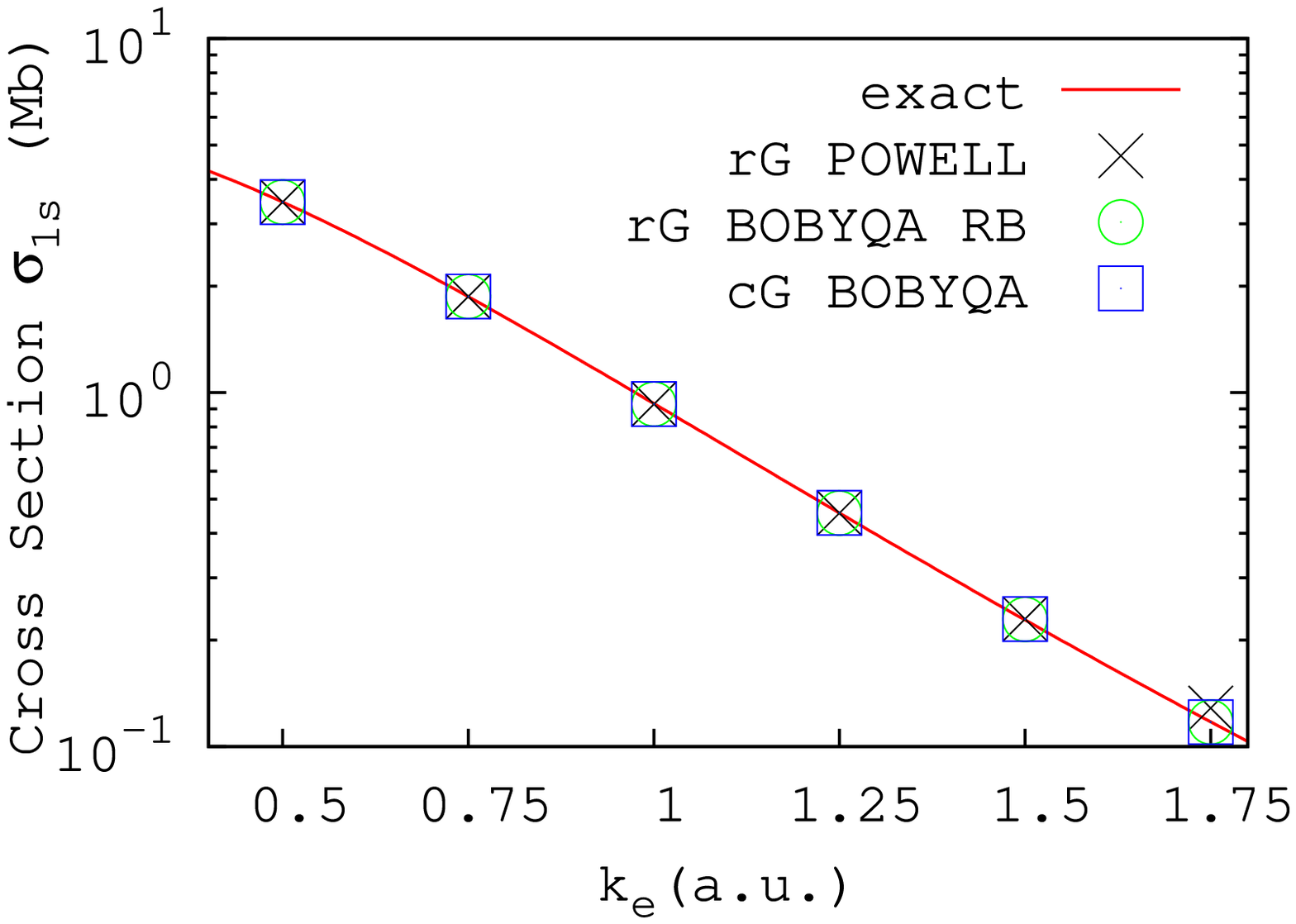}
			\end{minipage}
			\hfill
			\begin{minipage}{0.46\linewidth}
				\vspace{-0.35 cm}
				\includegraphics[width=\linewidth]{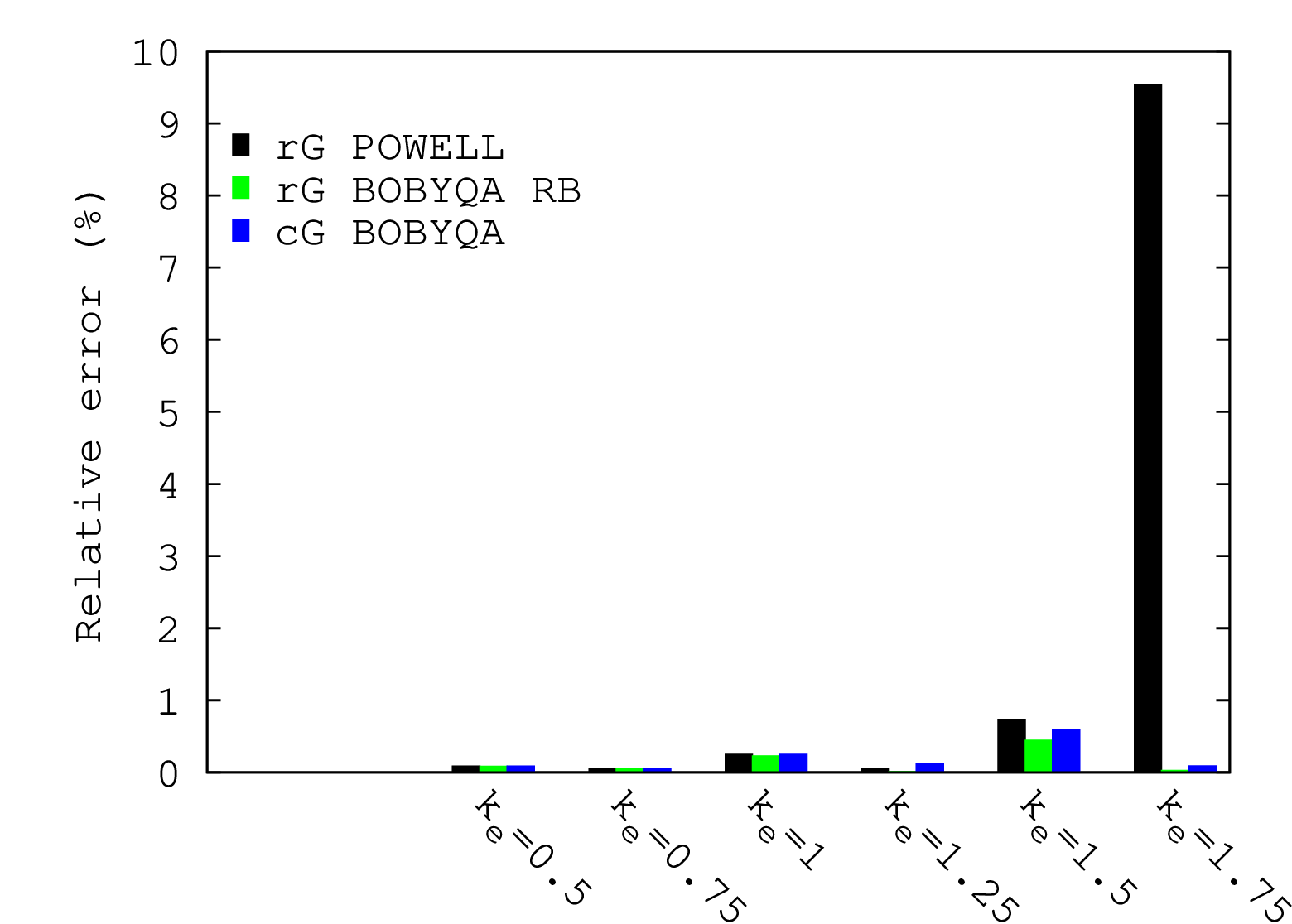}
			\end{minipage}
			\begin{minipage}{0.49\linewidth}
				\includegraphics[width=\linewidth]{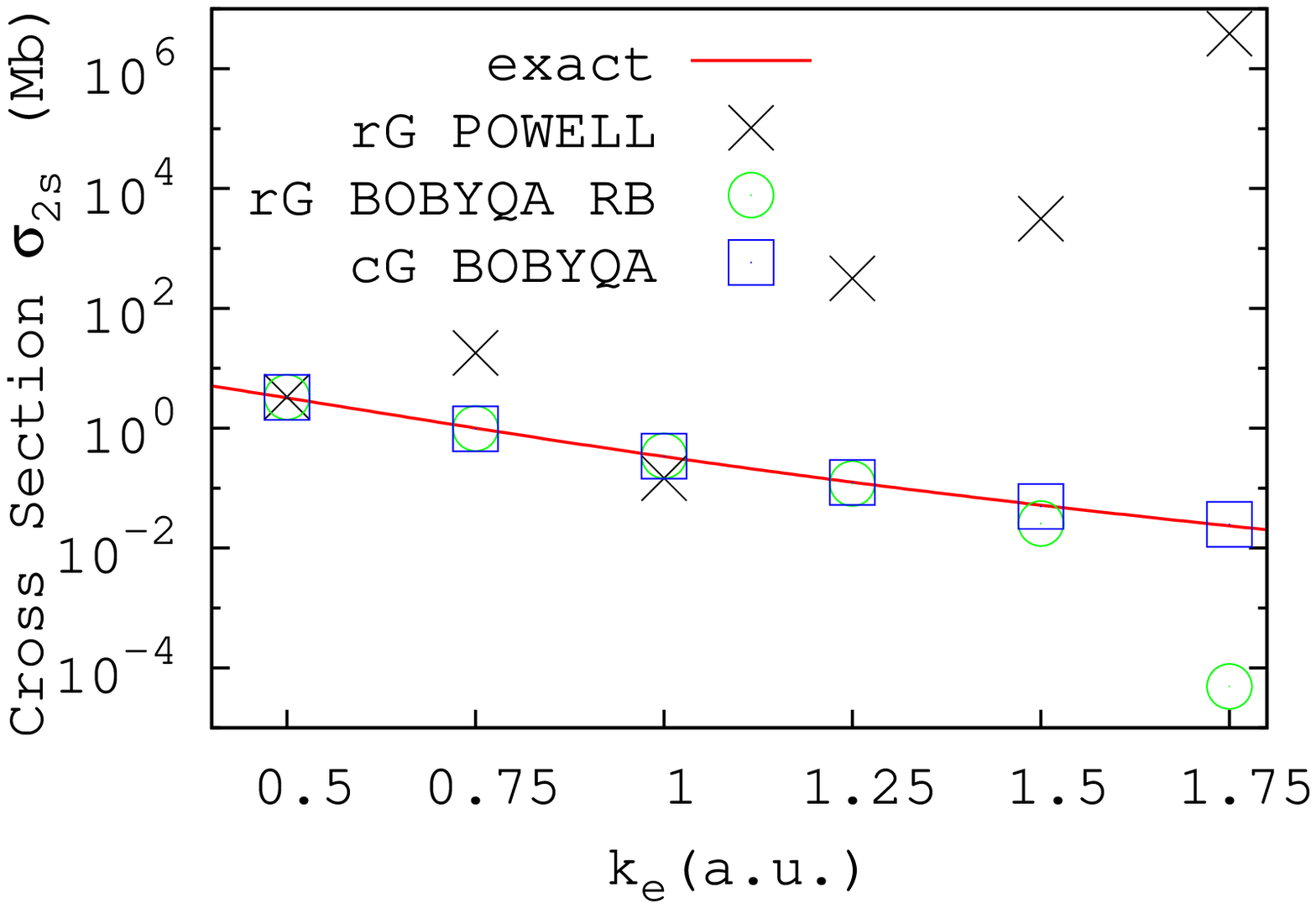}
			\end{minipage}
			\hfill
			\begin{minipage}{0.46\linewidth}
				\vspace{-0.35 cm}
				\includegraphics[width=\linewidth]{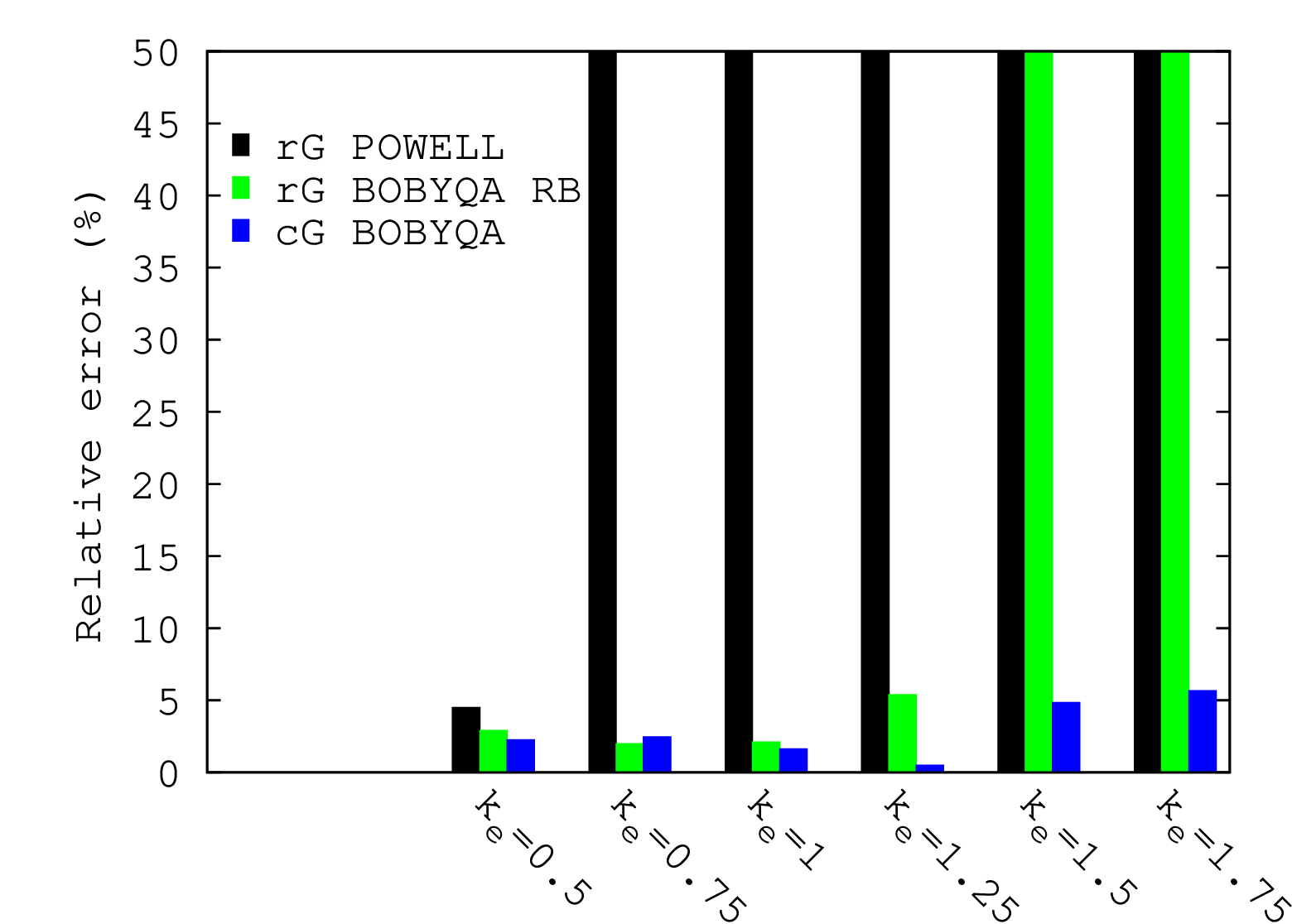}
			\end{minipage}
		\end{minipage}
		\vspace{-0.3 cm}
		\caption{\label{fig:PICS1s2s}Hydrogen photo-ionization cross section (in Mb) in terms of
			photo-electron wavenumber (in~$a.u.$). The hydrogen atom is initially in the $1s$
			state (upper panels), and in the $2s$ state (bottom panels). Black crosses, green circles
			and blue squares refer respectively to cross sections computed with real Gaussians
			with Powell, real Gaussians with BOBYQA by imposing reduced
			bounds, and complex Gaussians with BOBYQA. On the right panels, the histograms show the
			corresponding relative errors on the cross section computed with Gaussians.}
	\end{figure}

	Figure~\ref{fig:PICS1s2s}
	shows the photoionization cross section as a function of the photo-electron wavenumber
	$k_e$ for initial states $1s$ or $2s$, calculated with real Gaussians optimized using
	Powell (rG POWELL), real Gaussians optimized using BOBYQA with reduced bounds
	(rG BOBYQA RB), complex Gaussians optimized using BOBYQA (cG BOBYQA) and the exact results
	given by~\cite{harriman}:
	\begin{equation}
	\sigma_{1s} = \frac{2^5 \pi^2 e^{-4\frac{\arctan(k_e)}{k_e}}}
	{3cE_{\gamma}^4\left(1-e^{-2\frac{\pi}{k_e}}\right)} \text{,}
	\label{eq:PICS1s}
	\end{equation}
	and
	\begin{equation}
	\sigma_{2s} = \frac{\pi^2 \left( 8 + \frac{3}{E_{\gamma}} \right) e^{-4\frac{\arctan(2k_e)}{k_e}}}
	{6cE_{\gamma}^4\left(1-e^{-2\frac{\pi}{k_e}}\right)} \text{.}
	\label{eq:PICS2s}
	\end{equation}
	The corresponding relative errors are plotted in a histogram. The results are consistent
	with the quality of the fitting detailed in section~\ref{sec:level2}. Concerning H($1s$), there
	is no important difference between the different Gaussian basis sets except for the largest
	energy, $k_e=1.75$ where the rG POWELL method fails. As explained in section~\ref{sec:level2}, this
	is due to large deviations outside the fitting box. For H($2s$) the calculation
	with rG POWELL completely fails except for the small energy case $k_e=0.50$. The calculation
	with rG BOBYQA RB gives a small relative error at low energies, but fails above $k_e=1.25$. Only
	the cG BOBYQA choice achieves a very good accuracy in both $1s$ and $2s$ cases for all
	energies. These photoionisation cross section calculations demonstrate that care must be taken
	when using Gaussian expansions for continuum states, especially at higher energies and if they
	appear in matrix elements whose integrands extend on a large radial domain. For these delicate
	cases, complex Gaussians are clearly superior to their real counterpart.
	
	
	\section{\sffamily \Large  \label{sec:level6}Conclusions}
	
	In this work we applied a full nonlinear optimization to represent oscillating functions
	with both real and complex Gaussians by using a least square approach in combination with
	a quadratic approximation method. We have shown that real Gaussians may be sufficient to
	correctly span the space of functions within the fitting box, however they behave erroneously
	outside with sometimes very large errors due to diverging coefficients. Real Gaussian
	representation of continuum radial functions can nevertheless be used in ionization integrals
	if the decay factor associated with the bound state decreases fast enough to compensate
	this weakness. This may not be the case for highly excited states. Imposing a constraint
	on the Gaussian exponents reduces the error magnitude at small energies but it also
	reduces the fitting accuracy in the box and does not resolve the problem of
	ill-conditioned coefficients. On the other hand, complex Gaussians have an intrinsically
	oscillating behavior and are clearly more appropriate to fit bound functions that spread over
	large distances (highly excited or Rydberg states) and scattering functions.
	
	As an illustrative application, we have used the regular Coulomb
	wavefunctions fitted by Gaussians in hydrogen ionization problems.
	The accuracy of cross sections reproduced with complex Gaussians
	is as good as that obtained with real Gaussians at small energies
	while it is clearly superior at large energies, where real
	Gaussians fail completely. The application also allowed us to show
	the advantage provided by an all-Gaussian approach: the matrix
	elements can be evaluated analytically once all involved
	functions, bound and continuum, are expanded in real or complex
	Gaussians. In the present work we considered pure Coulomb
	continuum functions but our numerical strategy can be applied in
	the same manner to any one-center distorted wavefunctions
	generated analytically or numerically. We have already
	successfully tested the complex Gaussian representation of
	positive energy generalized Sturmian functions~\cite{GSF};
	with an adequately chosen asymptotic behavior, the latter can be
	used as an efficient basis to expand any distorted wave and thus
	to study collision problems.
	
	Extension to more complicated situations, including
	many-electron atoms and those molecules that can be described
	successfully with one-center expansions, do not present any
	additional technical difficulties. Going beyond scattering from a
	central potential, electron-electron integrals can be treated
	within a multipolar approach, and dealing with the ensuing
	integrals is part of our current investigations. Work is also
	ongoing to extend our proposal to study ionization processes with
	molecular targets, and thus to deal with multicenter integrals.
	For molecules the initial electronic wavefunctions
	are often already calculated in Gaussian bases. The representation
	of continuum states by Gaussians, even with complex exponents,
	should provide a way to simplify some of the multidimensional
	numerical integrations needed in scattering calculations. The
	use of the Gaussian product theorem will lead to some meaningless complex
	centers. These, however, do not lead to mathematical obstacles that cannot
	be dealt with as indicated in the work of Kuang and Lin~\cite{kuang1997,kuang1997b}.
	

	
	

	
	
	
	
	
	
	

	
	
	
	
	
	\clearpage
	
	

\begin{thebibliography}{99}
		
		%
		\bibitem{boys}
		Boys, S. F. \emph{Proc. R. Soc. London, Ser. A} \textbf{1950}, 200, 542-554.
		%
		\bibitem{Hill}
		Hill, J. G. \emph{Int. J. Quantum Chem.} \textbf{2013}, 113, 21-34.
		%
		\bibitem{Reeves}
		Reeves, C. M., Harrison, M. C. \emph{J. Chem. Phys.} \textbf{1963}, 39, 11-17.
		%
		\bibitem{Feller1979}
		Feller, D. F., Ruedenberg, K. \emph{Theor. Chim. acta} \textbf{1979}, 52, 231-251.
		%
		\bibitem{Huzinaga1}
		Huzinaga, S., Klobukowski, M. \emph{Chem. Phys. Lett.} \textbf{1985}, 120, 509-512.
		%
		\bibitem{Alexander1986}
		Alexander, S. A., Monkhorst, H. J., Szalewicz, K. \emph{J.
			Chem. Phys.} \textbf{1986}, 85, 5821-5825.
		%
		\bibitem{Petersson}
		Petersson, G. A., Zhong, S., Montgomery, J. A., Frisch, M. J. \emph{J. Chem. Phys.}
		\textbf{2003}, 118, 1101-1109.
		%
		\bibitem{Huzinaga1965}
		Huzinaga, S. \emph{J. Chem. Phys.} \textbf{1965}, 42, 1293-1302.
		%
		\bibitem{Reeves1965}
		Reeves, C. M., Fletcher, R. \emph{J. Chem. Phys.} \textbf{1965}, 42, 4073-4081.
		%
		\bibitem{oohata}
		O-ohata, K., Taketa, H., Huzinaga, S. \emph{J. Phys. Soc. Jpn.} \textbf{1966}, 21, 2306-2324.
		%
		\bibitem{Lim1966}
		Lim, T. K., Whitehead, M. A. \emph{J. Chem. Phys.} \textbf{1966}, 45, 4400-4413.
		%
		\bibitem{stewart1969}
		Stewart, R. F. \emph{J. Chem. Phys.} \textbf{1969}, 50, 2485-2495.
		%
		\bibitem{stewart1970}
		Stewart, R. F. \emph{J. Chem. Phys.} \textbf{1970}, 52, 431-438.
		%
		\bibitem{Harris}
		Harris, F. E. \emph{Rev. Mod. Phys.} \textbf{1963}, 35, 558-568.
		%
		\bibitem{Marante}
		Marante, C., Argenti, L., Mart\'{\i}n, F. \emph{Phys. Rev. A} \textbf{2014}, 90, 012506.
		%
		\bibitem{Cacelli1}
		Cacelli, I., Moccia, R., Rizzo, A. \emph{J. Chem. Phys.} \textbf{1993}, 98, 8742-8748.
		%
		\bibitem{Cacelli2}
		Cacelli, I., Moccia, R., Rizzo, A. \emph{J. Chem. Phys.} \textbf{1995}, 102, 7131-7141.
		%
		\bibitem{Cacelli3}
		Cacelli, I., Moccia, R., Rizzo, A. \emph{J. Chem. Phys.} \textbf{1998}, 57, 1895-1905.
		%
		\bibitem{Coccia1}
		Coccia, E., Mussard, B., Labeye, M., Caillat, J., Ta\"ieb, R., Toulouse, J., Luppi, E. \emph{Int.
			J. Quantum Chem.} \textbf{2016}, 116, 1120-1131.
		%
		\bibitem{Coccia2}
		Coccia, E., Luppi, E. \emph{Theor. Chem. Acc.} \textbf{2016}, 135, 43.
		%
		\bibitem{Coccia3}
		Coccia, E., Assaraf, R., Luppi, E., Toulouse, J. \emph{J. Chem. Phys.} \textbf{2017}, 147, 014106.
		%
		\bibitem{Rmatrix}
		Tennyson, J. \emph{Phys. Rep.} \textbf{2010}, 491, 29-76.
		%
		\bibitem{Kaufmann_1989}
		Kaufmann, K., Baumeister, W., Jungen, M. \emph{J. Phys. B - At. Mol. Opt.} \textbf{1989},
		22, 2223-2240.
		%
		\bibitem{nestmann}
		Nestmann, B. M., Peyerimhoff, S. D. \emph{J. Phys. B - At. Mol. Opt.} \textbf{1990}, 23, 773-777.
		%
		\bibitem{faure}
		Faure, A., Gorfinkiel, J. D., Morgan, L. A., Tennyson, J. \emph{Comput. Phys. Commun.}
		\textbf{2002}, 144, 224-241.
		%
		\bibitem{fiori}
		Fiori, M., Miraglia, J. E. \emph{Comput. Phys. Commun.} \textbf{2012}, 183, 2528-2534.
		%
		\bibitem{mccurdy1978}
		McCurdy Jr, C. W., Rescigno, T. N. \emph{Phys. Rev. Lett.} \textbf{1978}, 41, 1364-1368.
		%
		\bibitem{mccurdy1982}
		McCurdy, C. William, Mowrey, Richard C. \emph{Phys. Rev. A} \textbf{1982}, 25, 2529-2538.
		%
		\bibitem{isaacson1991}
		Isaacson, A. D. \emph{J. Chem. Phys.} \textbf{1991}, 94, 388-396.
		%
		\bibitem{white2017}
		White, A. F., Epifanovsky, E., McCurdy, C. W., Head-Gordon,
		M. \emph{J. Chem. Phys.} \textbf{2017}, 146, 234107.
		%
		\bibitem{Matsuzaki1}
		Matsuzaki, R., Yabushita, S. \emph{J. Comput. Chem.} \textbf{2017}, 38, 910-925.
		%
		\bibitem{Matsuzaki2}
		Matsuzaki, R., Yabushita, S. \emph{J. Comput. Chem.} \textbf{2017}, 38, 2030-2040.
		%
		\bibitem{Matsuzaki3}
		Matsuzaki, R., Asai, S., McCurdy, C. W., Yabushita, S. \emph{Theor. Chem. Acc.}
		\textbf{2014}, 133, 1-12.
		%
		\bibitem{kuang1997}
		Kuang, J., Lin, C. D. \emph{J. Phys. B. :  At. Mol. Opt. Phys.} \textbf{1997}, 30, 2529-2548.
		%
		\bibitem{kuang1997b}
		Kuang, J., Lin, C. D. \emph{J. Phys. B. :  At. Mol. Opt. Phys.} \textbf{1997}, 30, 2549-2567.
		%
		\bibitem{powell}
		Powell, M. J. D. \emph{Comput. J.} \textbf{1964}, 7, 155-162.
		%
		\bibitem{BOBYQA}
		Powell, M. J. D., Technical Report No. DAMTP 2009/NA06, Centre for Mathematical
		Sciences, University of Cambridge, UK, \textbf{2009}. The subroutine is available
		on https://www.zhangzk.net/
		and https://www.pdfo.net/
		%
		\bibitem{MathFunc}
		Abramowitz, M., Stegun, I., Eds; Handbook of Mathematical Functions; Dover, New York,
		\textbf{1964}; Chapter 14, p 538.
		%
		\bibitem{Dexpansion}
		Spielberger, L., Br\"auning, H., Muthig, A., Tang, J. Z., Wang, J., Qiu, Y., D\"orner, R., Jagutzki,
		O., Tschentscher, Th., Honkim\"aki, V., Mergel, V., Achler, M., Weber, Th., Khayyat, Kh.,
		Burgd\"orfer, J., McGuire, J., Schmidt-B\"ocking, H. \emph{Phys. Rev. A} \textbf{1999}, 59, 371-379.
		%
		\bibitem{gradshteyn2007}
		Gradshteyn, I. S., Ryzhik, I. M., Eds; Table of Integrals, Series, and
		Products; Academic Press, New York, \textbf{2007}; Chapter 6, p 706.
		%
		\bibitem{Dowell}
		McDowell, M. R. C., Coleman J. P., Eds; Introduction to the Theory of Ion-Atom
		collisions; North-Holland, New York, \textbf{1970}; Chapter 7, pp 322-324.
		%
		\bibitem{Burke}
		Burke, P. G.; R-Matrix Theory of Atomic Collisions; Springer, Berlin, \textbf{2011};
		Chapter 8, pp 380-390.
		%
		\bibitem{harriman}
		Harriman, J. M. \emph{Phys. Rev.} \textbf{1956}, 101, 594-598.
		%
		\bibitem{GSF}
		Gasaneo, G., Ancarani, L. U., Mitnik, D. M., Randazzo, J. M., Frapiccini, A. L.,
		Colavecchia, F. D. \emph{Adv. Quantum Chem.} \textbf{2013}, 67, 153-216.
		%
	\end{thebibliography}

\end{document}